\documentclass[11pt]{article}

\usepackage[latin1]{inputenc} 
\usepackage[nosort]{cite}
\usepackage[bulletsep]{collref}
\usepackage{graphicx}
\usepackage{color} 
\usepackage{bbm}
\usepackage{amsmath}
\usepackage{amssymb}
\usepackage{subfigure}
\usepackage{array}
\usepackage[bookmarks=true,hyperfigures=true]{hyperref}
\usepackage[a4paper,includeheadfoot,hmargin=2.5cm,vmargin=2cm,headheight=14pt,headsep=2.5ex,footskip=5.5ex]{geometry}

\providecommand{\hypersetup}[1]{}

\providecommand{\pdfbookmark}[3][]{}

\hypersetup{plainpages=false}
\hypersetup{pdfpagemode=UseOutlines}
\hypersetup{bookmarksnumbered=true}
\hypersetup{bookmarksopen=true}
\hypersetup{pdfstartview=FitH}
\hypersetup{colorlinks=false}
\hypersetup{citebordercolor={.6 .9 .6}}
\hypersetup{urlbordercolor={.7 .8 1}}
\hypersetup{linkbordercolor={1 .7 .7}}
\hypersetup{pdfborder={0 0 .5}}

\let\oldbfseries=\bfseries
\let\oldmdseries=\mdseries
\let\oldnormalfont=\normalfont
\renewcommand{\bfseries}{\oldbfseries\boldmath}
\renewcommand{\mdseries}{\oldmdseries\unboldmath}
\renewcommand{\normalfont}{\oldnormalfont\unboldmath}

\makeatletter
\let\old@makecaption=\@makecaption
\def\@makecaption{\small\old@makecaption}
\makeatother

\newcommand{\Smat}{\mathcal{S}}

\newcommand{\Action}{\mathcal{S}}
\newcommand{\Lagr}{\mathcal{L}}

\newcommand{\tr}{\mathop{\mathrm{tr}}}

\newcommand{\sign}{\mathop{\mathrm{sign}}}
\renewcommand{\Re}{\mathop{\mathrm{Re}}}
\renewcommand{\Im}{\mathop{\mathrm{Im}}}

\newcommand{\lt}{{\tilde \lambda}}

\newcommand{\order}{\mathcal{O}}

\newcommand{\Integers}{\mathbbm{Z}}

\newcommand{\Sphere}{S}  % {\mathbbm{S}}
\newcommand{\AdS}{\mathrm{AdS}}

\newcommand{\shift}{{\rm \bf D}}

\DeclareMathOperator{\csch}{csch}
\DeclareMathOperator{\sech}{sech}

\ifx\genfrac\sdflkaj

\else

\fi
\newcommand{\sfrac}[2]{{\textstyle\frac{#1}{#2}}}
\newcommand{\half}{\sfrac{1}{2}}
\newcommand{\ihalf}{\sfrac{i}{2}}

\newcommand{\Half}{\frac{1}{2}}
\newcommand{\iHalf}{\frac{i}{2}}

\newcommand{\alg}[1]{\mathfrak{#1}}
\newcommand{\grp}[1]{\mathrm{#1}}

\newcommand{\grSO}{\grp{SO}}

\newcommand{\algSU}{\alg{su}}

\newcommand{\algPSU}{\alg{psu}}

\newcommand{\lrbrk}[1]{\left(#1\right)}
\newcommand{\bigbrk}[1]{\bigl(#1\bigr)}
\newcommand{\Bigbrk}[1]{\Bigl(#1\Bigr)}

\newcommand{\lrsbrk}[1]{\left[#1\right]}
\newcommand{\bigsbrk}[1]{\bigl[#1\bigr]}
\newcommand{\Bigsbrk}[1]{\Bigl[#1\Bigr]}
\newcommand{\biggsbrk}[1]{\biggl[#1\biggr]}

\newcommand{\ket}[1]{\mathopen{|}#1\mathclose{\rangle}}
\newcommand{\bra}[1]{\mathopen{\langle}#1\mathclose{|}}
\newcommand{\braket}[2]{\mathopen{\langle}#1|#2\mathclose{\rangle}}

\newcommand{\lrabs}[1]{\left|#1\right|}

\newcommand{\comma}{\quad,\quad}

\newcommand{\nn}{\nonumber}
\newcommand{\nln}{\nonumber\\}
\newcommand{\nl}[1][0pt]{\nonumber\\[#1]&\hspace{-4\arraycolsep}&\mathord{}}

\newcommand{\earel}[1]{\mathrel{}&\hspace{-2\arraycolsep}#1\hspace{-2\arraycolsep}&\mathrel{}}
\newcommand{\eq}{\earel{=}}

\def\[{\begin{equation}}
\def\]{\end{equation}}
\def\<{\begin{eqnarray}}
\def\>{\end{eqnarray}}

\makeatletter
\def\mr@ignsp#1 {\ifx\:#1\@empty\else #1\expandafter\mr@ignsp\fi}%
\newcommand{\multiref}[1]{\begingroup%\let\protect\string%
\xdef\mr@no@sparg{\expandafter\mr@ignsp#1 \: }%
\def\mr@comma{}%
\@for\mr@refs:=\mr@no@sparg\do{\mr@comma\def\mr@comma{,}\ref{\mr@refs}}%
\endgroup}
\makeatother

\newcommand{\hypref}[2]{\ifx\href\asklfhas #2\else\href{#1}{#2}\fi}

\newcommand{\secref}[1]{Sec.~\multiref{#1}}

\newcommand{\appref}[1]{App.~\multiref{#1}}

\newcommand{\figref}[1]{Fig.~\multiref{#1}}
\renewcommand{\eqref}[1]{(\multiref{#1})}

\ifx\href\asklfhas\newcommand{\href}[2]{#2}\fi

\newcommand{\eps}{\varepsilon}

\let\vecarrow=\vec
\renewcommand{\vec}[1]{\mathbf{#1}}

\newcommand{\gt}{{\tilde \gamma}}

\newcommand{\Lc}{L_{\mathrm{c}}}
\newcommand{\Ls}{L_{\mathrm{s}}}
\newcommand{\lsg}{g}

\begin{document}

\begin{flushright}\footnotesize
%\texttt{arXiv:xxxx.xxxx}\\
\texttt{UUITP-09/13}\\
\texttt{TCDMATH 13-09}%
\end{flushright}
\vspace{1cm}

\begin{center}%
{\Large\textbf{\mathversion{bold}%
Comments on World-Sheet Form Factors in AdS/CFT
}\par}

\vspace{1.5cm}

\textrm{Thomas Klose$^{a}$ and Tristan McLoughlin$^{b}$} \vspace{8mm} \\
\textit{%
$^a$ Department of Physics and Astronomy, Uppsala University \\
SE-75108 Uppsala, Sweden \\
$^b$ School of Mathematics, Trinity College Dublin\\
College Green, Dublin 2, Ireland
} \\
\texttt{\\ thomas.klose@physics.uu.se, tristan@tcd.ie}

%%%%%%%%
\par\vspace{14mm}

\textbf{Abstract} \vspace{5mm}

\begin{minipage}{14cm}
We study form factors in the light-cone gauge world-sheet theory for strings in AdS$_5\times$S$^5$. We perturbatively calculate the two-particle form factor in a closed $\algSU(2)$ sector to one-loop in the near-plane-wave limit  and to two-loops in the Maldacena-Swanson limit. We also perturbatively solve the functional equation which follows from the form factor axioms for the world-sheet theory and show that the ``minimal" solution correctly reproduces the discontinuities of the perturbative calculations. Finally we propose a prescription, valid for polynomial orders of the inverse world-sheet length, for extracting the finite-volume world-sheet matrix element from the form factors and show that the two-excitation matrix element matches with the thermodynamic limit of  the spin-chain description of certain tree-level ${\cal N}=4$ SYM structure constants. 
\end{minipage}

\end{center}

\newpage

\tableofcontents

%%%%%%%%%%%%%%%%%%%%%%%%%%%%%%%%%%%%%%%%%%%%%%%%%%%%%%%%%%%%%%%%%%%%%%%%%%%
%%%%%%%%%%%%%%%%%%%%%%%%%%%%%%%%%%%%%%%%%%%%%%%%%%%%%%%%%%%%%%%%%%%%%%%%%%%
\section{Introduction}

Form factors serve as basic building blocks of observables in any quantum field theory and have played a particularly important role in the study of integrable models (see e.g. \cite{Smirnov:1992vz}). Recently they have been studied for the world-sheet theory of strings in $\AdS_5\times\Sphere^5$ \cite{Klose:2012ju}. Abstractly, they are matrix elements of local operators in the basis of asymptotic scattering states. As such they are both mathematically and conceptually very similar to the world-sheet S-matrix, and in particular, like the S-matrix, they are not directly related to their target-space counterparts. World-sheet form factors can rather be identified with matrix elements in the spin-chain model that is employed in the description of the dual gauge theory in the planar limit \footnote{This is by now an extensively studied subject that is reviewed in \cite{Arutyunov:2009ga, Beisert:2010jr} and where suitable references to the original literature can be found.}. This identification works in a similar way to that of the world-sheet and spin-chain S-matrix \cite{Staudacher:2004tk}.

\bigskip

Nevertheless, in principle world-sheet form factors can be used to construct target-space objects, which are then related to gauge theory quantities. In general, world-sheet form factors yield world-sheet correlation functions by expanding the latter as sums of products of the former. Then, world-sheet correlation functions of string vertex operators are nothing but gauge theory correlations functions of the dual operators. Expressing the gauge theory correlators in the language of spin-chains, as was done for tree-level three-point correlators  soon after the discovery of integrability of planar super Yang-Mills theory \cite{Okuyama:2004bd,Roiban:2004va,Alday:2005nd}, an even more direct link between world-sheet form factors and gauge theory quantities can be established. As was considered in \cite{Klose:2012ju}, and as we will discuss in more detail in \secref{sec:gauge-theory}, world-sheet form factors can be matched to the matrix elements of spin-chain operators in the strict thermodynamic limit and, by including finite-volume effects for the form factors, also at subleading orders. 

\bigskip

For large string tension, world-sheet form factors can be directly computed in the string sigma-model using perturbation theory. However, since the world-sheet theory is a two-dimen\-sional, integrable quantum field theory more efficient methods for determining form factors, which often lead to exact results, are known, see e.g.\ \cite{Smirnov:1992vz}. As part of the bootstrap program, the analytical properties of form factors are derived from general field theory considerations and then formulated as ``form factor axioms'' \cite{Weisz:1977ii,Karowski:1978vz,Smirnov:1992vz}. The idea is to construct functions that satisfy these axioms with the only direct reference to the underlying model being through the S-matrix and the spectrum of bound states. Perturbative calculations are then only necessary to fix the normalization or possibly to aid in identifying a given  solution with a specific form factor. In \cite{Klose:2012ju}, we investigated how the form factor axioms that are known for Lorentz-invariant models generalize to the non-relativistic world-sheet theory. We also checked the proposed properties against explicit world-sheet and spin-chain calculations and considered the weak-strong coupling interpolation for specific examples.

\bigskip

In this paper, we extend these considerations for the particular case of the two-particle form factor. In particular, we calculate explicitly the two-particle form factor
\< \label{eqn:YY-form-factor}
f(p_1,p_2) = \bra{0} \mathcal{O}(0) \ket{Y(p_1) Y(p_2)}
\>
for the quadratic operator
\< \label{eqn:YY-operator}
  \mathcal{O}(\vec{x}) = \Half :\!Y(\vec{x})^2\!:
\>
in a closed $\algSU(2)$ subsector of the string world-sheet theory. We do this to one-loop order for the full theory in the near-plane-wave limit \cite{Berenstein:2002jq} and to two-loop order in the truncated Maldacena-Swanson or near-flat theory \cite{Maldacena:2006rv}. These explicit results provide useful data regarding the structure of the form factors and provide further checks of the world-sheet axioms proposed in \cite{Klose:2012ju}.

\bigskip

While finding solutions of the axioms is a promising method for finding exact, all-order in $\lambda$, form factors, due to the complicated nature of the world-sheet S-matrix, particularly the dressing phase \cite{Beisert:2006ib, Beisert:2006ez}, the answers may be involved and it is useful to start with simple limits. In this work we solve the proposed axioms perturbatively for the two-particle form factor of fundamental fields in the closed $\algSU(2)$ subsector and compare these results with our explicit perturbative calculations. While we focus on the two-particle form factor, which is the simplest non-trivial case, in many regards this acts a fundamental building block for higher point cases.

\bigskip

In general, the full two-particle form factor is a product of three components: the normalization, a factor providing the appropriate bound state poles, and a ``minimal" solution. The minimal solution is a solution to Watson's equations---i.e.\ to the periodicity and the permutation axioms for the case of two external particles---without poles in the physical region. It is part of all form factors but generically is not by itself the form factor for any operator. A formal expression for the solution of Watson's equations is as an infinite product of two-particle S-matrices with shifted relative rapidity. One way of making this formal expression precise is to write the S-matrix in a suitable integral representation and then to carry out the infinite product, for example this is known to yield the correct results for breathers in sine-Gordon theory. In this paper, we  apply this formula to the world-sheet theory. However, as we do not have an appropriate integral expression for the \emph{exact} world-sheet S-matrix, we need to work perturbatively. Thus, we consider the near-plane-wave theory to first order and the near-flat-space theory to second order, where the corresponding S-matrices of \cite{Arutyunov:2004vx} and \cite{Maldacena:2006rv} (see \cite{Klose:2007rz} for a perturbative calculation), respectively, are simple enough to be re-written in an appropriate form. 

\bigskip

We can also  see how this minimal solution is contained in actual form factors. We find that the minimal solution correctly captures the terms in the form factor that have non-rational dependence on the particle momenta. This is fully in line with expectations, as the form factor axioms precisely describe the discontinuities of the form factors under analytic continuation. The rational terms will be provided by an independent factor that multiplies the minimal solution. This additional factor should be fixed, or at least constrained,  by imposing conditions on the poles that occur when the momenta are such that the external particles can form bound states or  that internal particles go on-shell.

\bigskip

In the thermodynamic limit, corresponding to infinite charges and infinite world-sheet volume, a direct comparison can be made between the results of the near-plane-wave string theory and the spin-chain calculations describing the tree-level structure constants. This match at low orders in the gauge theory perturbative expansion is well known for the spectrum of anomalous dimensions and it is expected that it will fail at sufficiently high loop-order. Nonetheless it is useful to pursue this serendipitous matching  for the insight it provides into using world-sheet form factors to calculate gauge theory structure constants. A key step in going beyond the strict thermodynamic limit is to consider form factors in finite volume: here we propose that by considering external momenta that satisfy the string Bethe ansatz equations and including a density of states factor that one captures all polynomial corrections in the world-sheet length, $\Ls$. Furthermore, we show that the world-sheet $1/\sqrt{\lambda}$ corrections reproduce the finite spin-chain length, $\Lc$, corrections, at least where reliable comparison can be made.

%%%%%%%%%%%%%%%%%%%%%%%%%%%%%%%%%%%%%%%%%%%%%%%%%%%%%%%%%%%%%%%%%%%%%%%%%%%
%%%%%%%%%%%%%%%%%%%%%%%%%%%%%%%%%%%%%%%%%%%%%%%%%%%%%%%%%%%%%%%%%%%%%%%%%%%
\section{Perturbative World-Sheet Theory Computations}

\begin{figure}
\begin{center}
\subfigure[Tree level]{\label{fig:F2-feynman-tree}
                       \includegraphics[scale=0.7]{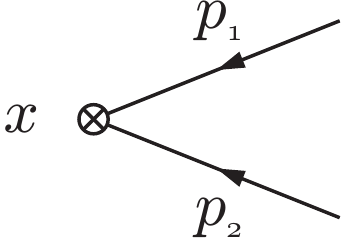}}
\hspace{30mm}
\subfigure[One loop]{\label{fig:F2-feynman-one}
                     \includegraphics[scale=0.7]{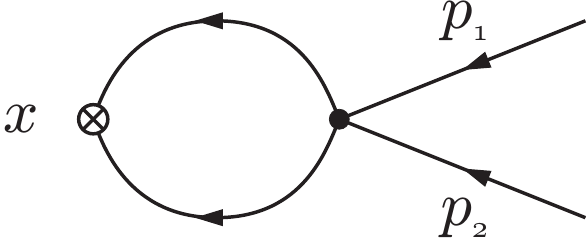}} \\[4mm]
\subfigure[Two loop]{\label{fig:F2-feynman-two}
                     \includegraphics[scale=0.7]{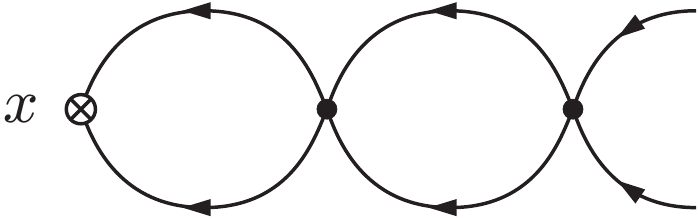}\hspace{10mm}
                     \includegraphics[scale=0.7]{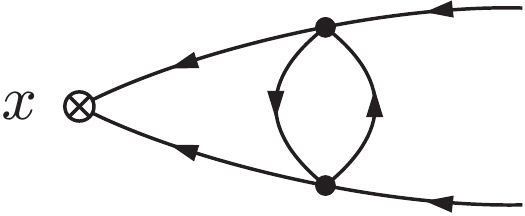}\hspace{10mm}
                     \includegraphics[scale=0.7]{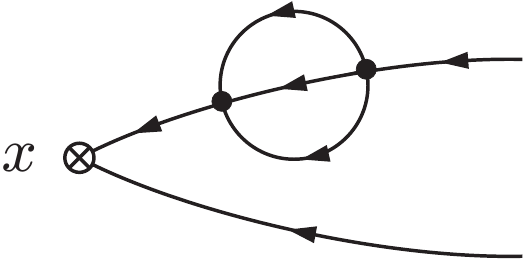}}
\end{center}
\caption{\textbf{Feynman diagrams for two-particle form factors.}}
\label{fig:F2-feynman}
\end{figure}

In this section, we present the perturbative computations of the two-particle form factor \eqref{eqn:YY-form-factor} in the world-sheet theory for strings in $\AdS_5\times\Sphere^5$. The field
\<
  Y = \frac{1}{\sqrt{2}} (Y_1 + i Y_2)
\>
is a complex combination of two scalar fields on $\Sphere^5$. Firstly, we compute the form factor in the near-BMN or near-plane-wave limit of the world-sheet theory to one-loop order in the world-sheet coupling constant $\lambda^{-1/2}$. This calculation is in many regards similar to the perturbative calculation of
the world-sheet S-matrix \cite{Klose:2006zd} from which many notations and details will be taken. Secondly, we will extend the one-loop computation of \cite{Klose:2012ju} of this form-factor in the near-flat-space limit to two loops. The relevant Feynman diagrams are given in \figref{fig:F2-feynman}.

%%%%%%%%%%%%%%%%%%%%%%%%%%%%%%%%%%%%%%%%%%%%%%%%%%%%%%%%%%%%%%%%%%%%%%%%%%%
%%%%%%%%%%%%%%%%%%%%%%%%%%%%%%%%%%%%%%%%%%%%%%%%%%%%%%%%%%%%%%%%%%%%%%%%%%%
\subsection{Near-Plane-Wave World-Sheet Theory}
\label{sec:npw-pert}
For the perturbative calculation, we start from the light-cone gauge-fixed Lagrangian for the complex scalar $Y$ and its conjugate $\bar{Y}$ to quartic order in the fields. The quadratic part is simply that of a massive relativistic particle
\<
  \Lagr_2 = \partial Y \partial \bar{Y} - Y \bar{Y}~.
\>
The quartic terms depend explicitly on the gauge choice and in general $a$-gauge \cite{Arutyunov:2006gs} they are given by 
\<
  \Lagr_4 = 2 Y \acute{Y} \bar{Y} \acute{\bar{Y}} + \frac{1-2a}{2} \Bigbrk{ (\partial Y)^2 (\partial \bar{Y})^2 -  Y^2 \bar{Y}^2 }~.
\>
The Lagrangian is normalized such that the action is given by
\< \label{eqn:npw-action}
  \Action = \frac{\sqrt{\lambda}}{2\pi}\int d^2\sigma~\Lagr
  \qquad\text{where}\qquad
  \Lagr= \Lagr_2+ \Lagr_4+\dots~, 
\>
and in the calculation of world-sheet form factors we are considering the decompactified theory defined on the infinite plane such that there are
well defined asymptotic scattering states. For such asymptotic particle states, it is useful to introduce the usual rapidity parameters
\< \label{eqn:usual-rapitidies}
  \epsilon_i = \cosh \theta_i
	\qquad \text{and}\qquad
  p_i = \sinh\theta_i
\>
and the combinations $\theta=\theta_2-\theta_1$ and $\tilde\theta=\theta_2+\theta_1$. In a Lorentz invariant theory only the first of these would appear in the S-matrix or the form factors but for the world-sheet theory both are necessary.

\bigskip

At tree-level, the form factor is given by the product of the bare wave-functions, $Z(\epsilon) = 1/\sqrt{2\epsilon}$, of the external states, i.e.\
\<  \label{eqn:ff-npw-tree}
f^{(0)}(p_1, p_2)=\frac{1}{2\sqrt{\epsilon_1\epsilon_2}}~,
\>
which explains the choice of normalization in \eqref{eqn:YY-operator}. At one-loop order, the form factor is given by the following bubble integral with non-trivial numerator factors
\<
  f^{(1)}(p_1,p_2) = 2 \frac{i\pi}{\sqrt{\lambda}} \frac{1}{2\sqrt{\epsilon_1\epsilon_2}} \int\!\frac{d^2k}{(2\pi)^2} \frac{
	  (p_1+p_2)^2 + (1-2a)\bigsbrk{ \vec{p}_1\cdot\vec{p}_2 \,\, \vec{k}\cdot (\vec{p}_1+\vec{p}_2-\vec{k}) - 1 }
	}{\bigsbrk{\vec{k}^2 - 1}\bigsbrk{(\vec{p}_1+\vec{p}_2-\vec{k})^2 - 1}} ~.
\>
The integral over the term containing $\vec{k}^2$ in the numerator is divergent in two dimensions. In the $a=1/2$ gauge, this term vanishes and the result is finite. However, we expect to obtain a (off-shell) match with the thermodynamic limit of the spin-chain only for $a=1$ as seen previously in \cite{Klose:2012ju}, though in that case at tree-level. Thus, we regularize the integral by going to $d=2-2\eps$ dimensions. We find
\<
  f^{(1)}(p_1,p_2) = 2 \frac{i\pi}{\sqrt{\lambda}} \frac{1}{2\sqrt{\epsilon_1\epsilon_2}} \Bigsbrk{
	  (p_1+p_2)^2 B(p_1,p_2) + (1-2a) \bigbrk{ \vec{p}_1\cdot\vec{p}_2 \, X(p_1,p_2) - B(p_1,p_2) } }
\>
where $X(p_1,p_2)$ is most easily expressed in terms of the relative rapidity as
\<
  X(p_1,p_2) = \frac{i}{4\pi} \lrsbrk{ \frac{1}{\eps} + \ln 4\pi - \gamma_E - (\theta-i\pi\sign\theta) \coth\theta + \order(\eps) } ~.
\>
Note that $\sign\theta$ ensures that $X(p_1,p_2)$ is symmetric under the exchange of the momenta. In this notation, the bubble integral is
\<
  B(p_1,p_2) = -\frac{i}{4\pi} (\theta-i\pi\sign\theta) \csch\theta ~.
\>
The function $X$ contains a divergence in $d=2$ which can be removed by renormalizing the composite operator.  We choose a $\overline{\rm MS}$-scheme (reviewed in \appref{app:Opren}) which effectively boils down to dropping the ``$1/\eps + \mathrm{const.}$'' terms. Labelling the renormalized result $f^{(1)}_{\mathrm{ren}}$, assuming $\theta>0$, and simplifying we find
\< \label{eqn:ff-ren-rapid}
  f^{(1)}_{\mathrm{ren}}(\theta_1,\theta_2) = \frac{1}{\sqrt{\lambda}} \frac{1}{2\sqrt{\epsilon_1\epsilon_2}} (\theta-i\pi)
	\lrsbrk{ \coth\frac{\theta}{2} \sinh^2\frac{\tilde{\theta}}{2} + \frac{1-2a}{2} \sinh\theta } ~.
\>

As we will see, to compare this result to the perturbative solution of Watson's equation, \eqref{eqn:ff-npw-Watson},  we will need to set $a=1/2$. This is simply due to the technical fact that when solving the functional equations, we  need to write the S-matrix as an integral expression and found such an expression only for $a=1/2$. In this gauge, the form factor to one-loop order is
\< \label{eqn:ff-ren-rapid-a12}
  f(\theta_1,\theta_2) = \frac{1}{2\sqrt{\epsilon_1\epsilon_2}} \lrsbrk{ 1 + \frac{1}{\sqrt{\lambda}} (\theta-i\pi)
	\coth\frac{\theta}{2} \sinh^2\frac{\tilde{\theta}}{2} } ~.
\>
While this calculation is one-loop, it is not sensitive to any of the fields outside the $\algSU(2)$ sector. This makes the calculation particularly straightforward, however, it does not provide a particularly stringent test of the form factor axioms for the world-sheet theory. Perturbative calculations at two-loops would generically involve all the additional bosonic fields, including those from the AdS space, and the fermions. To date, there has not been a full one-loop perturbative calculation of the world-sheet S-matrix (which would be analogous to the two-loop form factor) except in the near-flat limit \cite{Maldacena:2006rv} which, at least to the orders checked \cite{Klose:2007wq,Klose:2007rz, Puletti:2007hq}, is a consistent truncation of the full theory.

%%%%%%%%%%%%%%%%%%%%%%%%%%%%%%%%%%%%%%%%%%%%%%%%%%%%%%%%%%%%%%%%%%%%%%%%%%%
%%%%%%%%%%%%%%%%%%%%%%%%%%%%%%%%%%%%%%%%%%%%%%%%%%%%%%%%%%%%%%%%%%%%%%%%%%%
\subsection{Near-Flat World-Sheet Theory}

The world-sheet Lagrangian in the near-flat limit  can be written as \cite{Klose:2007wq,Klose:2007rz}
\< \label{eqn:MS-action}
 \Lagr \eq
  \tfrac{1}{2}(\partial \vecarrow{Y})^2-\tfrac{1}{2}\,\vecarrow{Y}^2
 +\tfrac{1}{2}(\partial \vecarrow{Z})^2-\tfrac{1}{2}\,\vecarrow{Z}^2
 +\tfrac{i}{2}\psi \tfrac{\partial^2+1}{\partial_-}\,\psi \nl[1mm]
 + \gamma\,(\vecarrow{Y}^2-\vecarrow{Z}^2)\bigbrk{(\partial_- \vecarrow{Y})^2+(\partial_- \vecarrow{Z})^2}
 +i\gamma\,(\vecarrow{Y}^2-\vecarrow{Z}^2)\psi\partial_-\psi \nl[1mm]
 +i\gamma\,\psi\bigbrk{\partial_- Y_{i'} \Gamma_{i'} + \partial_- Z_i \Gamma_i}
           \bigbrk{Y_{j'} \Gamma_{j'} - Z_j \Gamma_j}\psi \nl[1mm]
 -\tfrac{\gamma}{24}\bigbrk{\psi\Gamma_{i'j'}\psi\,\psi\Gamma_{i'j'}\psi
                      -\psi\Gamma_{ij}  \psi\,\psi\Gamma_{ij}  \psi} ~,
\>
where $\vecarrow{Y}$ and $\vecarrow{Z}$ are the $4+4$ bosonic fields transverse to the light-cone in $\Sphere^5$ and $\AdS_5$, respectively. The eight fermionic degrees of freedom are comprised in an $\grSO(8)$ Majorana-Weyl spinor $\psi$. The prefactor $\sqrt{\lambda}/(2\pi)$ that was present in \eqref{eqn:npw-action} has been removed by a rescaling of the fields and is now present as $\gamma = \pi/\sqrt{\lambda}$ in front of the interaction terms.

We quantize the theory with $\sigma^+ = \tau + \sigma$ considered as world-sheet time. Correspondingly, $p_+ = \half(\epsilon+p)$ should be interpreted as the energy of the particle and $p_- = \half(\epsilon - p)$ as its momentum. For convenience, we introduce the shorthand notation
\<
  \xi \equiv p_+ \qquad \text{and} \qquad \eta \equiv p_- \; .
\>
For further details on the derivation and the quantization of the model, we refer to our one-loop form factor computation \cite{Klose:2012ju}, 
to the perturbative S-matrix calculations \cite{Klose:2007wq,Klose:2007rz, Puletti:2007hq} and, of course, to the original work \cite{Maldacena:2006rv}.

\bigskip

The Feynman diagrams up to two-loops have the structures drawn in \figref{fig:F2-feynman}. The tree-level and one-loop results were obtained in \cite{Klose:2012ju} and read
\< \label{eqn:nfs-tree-and-one}
  f^{(0)}(\eta_1,\eta_2) = \frac{1}{2\sqrt{\eta_1\eta_2}}
  \comma
  f^{(1)}(\eta_1,\eta_2) = \frac{-i\gamma}{\sqrt{\eta_1\eta_2}} \eta_{12}^2 B(\eta_1,\eta_2) ~,
\>
where the multi-index notation means $\eta_{ij...} = \eta_i + \eta_j + \ldots$. In these variables, the bubble integral is (for $\eta_1>\eta_2>0$)
\< \label{eqn:bubble}
  B(\eta_1,\eta_2) = 
  \frac{i}{2\pi} \frac{\eta_1 \eta_2}{\eta_{12} \eta_{1\bar{2}}}
  \lrsbrk{ \ln\lrbrk{\frac{\eta_2}{\eta_1}} + i\pi }~,
\>
where the bar in $\eta_{1\bar{2}}$ is defined to mean that $\eta_2$ is subtracted from $\eta_1$ rather than added to it.

The two-loop diagrams are the double-bubble (``db''), the wineglass (``wg''), and the sunset diagram (``ss''), see \figref{fig:F2-feynman-two}. Summing the contributions from the various vertices gives
\<
  \label{eqn:ff2-two-bubb}
  f^{(2,\mathrm{db})}(\eta_1,\eta_2) \eq
    - \frac{2\gamma^2}{\sqrt{\eta_1\eta_2}} \eta_{12}^4 B(\eta_{12})^2 ~, \\
  \label{eqn:ff2-two-wine}
  f^{(2,\mathrm{wg})}(\eta_1,\eta_2) \eq
    - \frac{8\gamma^2}{\sqrt{\eta_1\eta_2}}
      \Bigsbrk{ 4 \eta_1^2 \eta_2^2 W_0(\eta_1,\eta_2)
              + 8 \eta_1 \eta_2 \eta_{1\bar{2}} W_1(\eta_1,\eta_2) \nl \hspace{20mm}
              + (\eta_1^2 - 6 \eta_1 \eta_2 + \eta_2^2 ) W_2(\eta_1,\eta_2) } ~, \\
  \label{eqn:ff2-two-sun}
  f^{(2,\mathrm{ss})}(\eta_1,\eta_2) \eq
      -\frac{\gamma^2}{4\sqrt{\eta_1\eta_2}} \lrbrk{\frac{1}{\pi^2} - \frac{1}{12}} (\eta_1^4 + \eta_2^4) ~.
\>
The contribution from the sunset diagram is nothing but the two-loop renormalization of the wave-function given by \cite{Klose:2007rz}
\<
  Z(\eta) = \frac{1}{2\eta} \lrsbrk{ 1 - \gamma^2 \lrbrk{\frac{1}{\pi^2} - \frac{1}{12}} \eta^4 + \order(\gamma^3) } ~.
\>
The three types of wineglass integrals, $W_0 \equiv W_{0000}$, $W_1 \equiv W_{1000}$, and $W_2 \equiv W_{1100}+W_{2000}$, are special cases of\footnote{This is related to the wineglasses, $\tilde{W}_{rstu}(p,p')$ used in \cite{Klose:2007rz} by $W_{rstu}(\eta_1,\eta_2) = (-1)^{t+u} \tilde{W}_{rstu}(\eta_1,-\eta_2)$.}
\< \label{eqn:wineglass}
  W_{rstu}(\eta_1,\eta_2) = \int \frac{d^2k\,d^2q}{(2\pi)^4} \frac{\eta_k^r \eta_q^s (\eta_1-\eta_k-\eta_q)^t (\eta_2+\eta_k+\eta_q)^u}{[\vec{k}^2 - 1][\vec{q}^2 - 1][(\vec{p}_1-\vec{k}-\vec{q})^2 - 1][(\vec{p}_2+\vec{k}+\vec{q})^2 - 1]} ~.
\>
The relevant integrals were previously evaluated in the two-loop computation of the world-sheet S-matrix \cite{Klose:2007rz}. Expressing them in terms of the bubble \eqref{eqn:bubble}, we find
\<
  W_0(\eta_1,\eta_2) \eq \frac{\eta_1\eta_2}{16\pi^2} \lrsbrk{ \frac{\pi^2}{4\eta_{1\bar{2}}^2} + \frac{2\pi i}{\eta_1\eta_2} B(\eta_1,\eta_2) - \frac{4\pi^2}{\eta_1\eta_2} B(\eta_1,\eta_2)^2 } ~, \\
  W_1(\eta_1,\eta_2) \eq \frac{\eta_1\eta_2}{16\pi^2} \lrsbrk{ \frac{\pi^2}{8\eta_{1\bar{2}}} - 2\pi^2 \frac{\eta_{1\bar{2}}}{\eta_1\eta_2} B(\eta_1,\eta_2)^2 } ~, \\
  W_2(\eta_1,\eta_2) \eq \frac{\eta_1\eta_2}{16\pi^2} \lrsbrk{ \frac{\pi^2}{12} - 2\pi^2 \frac{\eta_{1\bar{2}}^2}{\eta_1\eta_2} B(\eta_1,\eta_2)^2 } ~.
\>
Now, we can write the sum of all two-loop contributions, \eqref{eqn:ff2-two-bubb}, \eqref{eqn:ff2-two-wine}, and \eqref{eqn:ff2-two-sun}, as
\<
\label{eqn:nfs-two}
  f^{(2)}(\eta_1,\eta_2) \eq
     - \frac{\gamma^2}{4\sqrt{\eta_1\eta_2}} \biggsbrk{
          8 \eta_{12}^4 B(\eta_{12})^2 
        + \frac{\eta_1 \eta_2 (\eta_1^2+\eta_2^2) (\eta_1^2 + 4 \eta_1\eta_2 + \eta_2^2)}{6\eta_{1\bar{2}}^2} \nl \hspace{25mm}
        + \frac{16i}{\pi} \eta_1^2\eta_2^2 B(\eta_{12})
        - 4 ( \eta_1^4 + 6 \eta_1^2\eta_2^2 + \eta_2^4 ) B(\eta_{12})^2 \nl \hspace{25mm}
        + \lrbrk{\frac{1}{\pi^2} - \frac{1}{12}} (\eta_1^4 + \eta_2^4)
    } ~.
\>
Although the computation is rather similar to the computation of the two-particle world-sheet S-matrix, the final expression for the two-particle form factor is much more complicated; in particular in contains terms $\sim(\log)^2$. The difference can be traced back to the fact that the two-particle out-state in the S-matrix computation carries the sum of two on-shell momenta, while the operator in the form factor computation can absorb an off-shell momentum.

We can check that the above result satisfies the permutation property by computing
$\Delta f \equiv f(\eta_2,\eta_1) - f(\eta_1,\eta_2)$. Note, however, that this test is sensitive only to the bubble terms. Using $B(\eta_2,\eta_1) = B(\eta_1,\eta_2) + \eta_1\eta_2/\eta_{12}\eta_{1\bar{2}}$, we find 
\<
  (\Delta f)^{(2)}(\eta_1,\eta_2) \eq -\frac{\gamma^2}{\sqrt{\eta_1\eta_2}} \biggsbrk{
\frac{ \eta_1 \eta_2 \left(\eta_1^4+8 \eta_1^3 \eta_2+6 \eta_1^2 \eta_2^2+8 \eta_1 \eta_2^3+\eta_2^4\right) \left(\eta_1 \eta_2+2 \eta_{12} \eta_{1\bar{2}} B(\eta_1,\eta_2)\right)}{\eta_{12}^2 \eta_{1\bar{2}}^2} \nl \hspace{20mm}
+\frac{4 i \eta_1^3 \eta_2^3}{\eta_{12} \eta_{1\bar{2}} \pi}
} ~,
\>
or with the explicit expression for the bubble
\<
  (\Delta f)^{(2)} = -\frac{4i\gamma^2}{\pi\sqrt{\eta_1\eta_2}} \frac{\eta_1^2\eta_2^2}{\eta_{12}\eta_{1\bar{2}}} \lrsbrk{ \eta_1\eta_2 + \frac{\eta_1\eta_2(\eta_1^2 + \eta_2^2) + \eta_{12}^4/4}{\eta_{12}\eta_{1\bar{2}}} \ln\frac{\eta_2}{\eta_1}} ~.
\>
It is straightforward to verify that this matches the prediction of the permutation property given by
\< \label{eqn:permutation-property}
  (\Delta f)^{(2)} = f^{(1)}(\eta_1,\eta_2) \Smat^{(0)}(\eta_1,\eta_2) + f^{(0)}(\eta_1,\eta_2) \Smat^{(1)}(\eta_1,\eta_2)
\>
with $f^{(0)}(\eta_1,\eta_2)$ and $f^{(1)}(\eta_1,\eta_2)$ from \eqref{eqn:nfs-tree-and-one} and the zeroth and first order of the S-matrix \cite{Klose:2007wq}
\<
 \Smat^{(0)}(\eta_1,\eta_2) \eq -2i\gamma \eta_1\eta_2 \frac{\eta_{12}}{\eta_{1\bar{2}}} ~, \\
 \Smat^{(1)}(\eta_1,\eta_2) \eq -2\gamma^2 \eta_1^2 \eta_2^2 \frac{\eta_{12}^2}{\eta_{1\bar{2}}^2}
                              - \frac{8i\gamma^2}{\pi} \frac{\eta_1^3\eta_2^3}{\eta_{12}\eta_{1\bar{2}}} \lrbrk{ 1 + \frac{\eta_1^2+\eta_2^2}{\eta_{12}\eta_{1\bar{2}}} \ln\frac{\eta_2}{\eta_1}} ~.
\>
We note that the presence of the $(\log)^2$-terms is essential for the two-loop permutation property to hold. While a single log-factor has a constant discontinuity, log-squared has a discontinuity that is proportional to the logarithm of the momenta. Such a discontinuity in the form factor is a prerequisite for being able to match the one-loop S-matrix as required by the permutation property \eqref{eqn:permutation-property}.

%%%%%%%%%%%%%%%%%%%%%%%%%%%%%%%%%%%%%%%%%%%%%%%%%%%%%%%%%%%%%%%%%%%%%%%%%%%
%%%%%%%%%%%%%%%%%%%%%%%%%%%%%%%%%%%%%%%%%%%%%%%%%%%%%%%%%%%%%%%%%%%%%%%%%%%
\section{Solutions of the Functional Equations}

One method for deriving form factors in an integrable model involves solving generalised Watson equations involving the exact S-matrix. These equations encode various properties expected in a sensible quantum field theory such as unitarity, crossing symmetry (properly understood) and factorisation of the S-matrix. The latter is a particularly powerful property in an integrable theory with an infinite number of conserved quantities and can be formalised in terms of the Zamolodchikov algebra, see e.g. \cite{Smirnov:1992vz}. Form factors can then be built from solutions of these functional equations with appropriate analytical properties. In general, the functional equations are matrix valued with intricate group structure, however, we will focus on the simplest case: two-particle form factors in a rank-one sector. 

%%%%%%%%%%%%%%%%%%%%%%%%%%%%%%%%%%%%%%%%%%%%%%%%%%%%%%%%%%%%%%%%%%%%%%%%%%%
%%%%%%%%%%%%%%%%%%%%%%%%%%%%%%%%%%%%%%%%%%%%%%%%%%%%%%%%%%%%%%%%%%%%%%%%%%%
\subsection{Formal Solution of the Functional Equations}

We first review the well known relativistic case where the  S-matrix and the two-particle form factor only depend on the difference of the external particle rapidities, $\theta=\theta_2-\theta_1$. The two-particle functional equations in a rank-one sector are
\< \label{eqn:Watson}
  f(\theta) = f(-\theta) \Smat(\theta) \qquad\text{and}\qquad f(i \pi-\theta)=f(i \pi+\theta)~.
\>
The first equation is self-consistent only if the S-matrix satisfies $\Smat^{-1}(\theta)=\Smat(-\theta)$. Combining these two equations, we obtain
\< \label{eqn:Watson-combined}
  f(\theta+2 i \pi) = f(\theta) \Smat(-\theta)~.
\>
In solving this equation, see e.g.\ \cite{Karowski:1978vz}, it is assumed that $f(\theta)$ is meromorphic in the physical strip $0\leq \Im\theta\leq \pi$ with poles only on the imaginary axis
\footnote{It is perhaps more natural to define the form factor on the double cover $0\leq \Im\theta\leq 2 \pi$
e.g. \cite{Smirnov:1992vz}. Indeed the contour argument \cite{Karowski:1978vz} used to determine the two-particle minimal solution uses the fact that it is analytic with no zeroes or poles in this larger space.}. With appropriate asymptotic conditions, the two particle form factor can be written as
\<
f(\theta)=k(\theta)f_{\rm min}(\theta)~,
\>
where $f_{\rm min}(\theta)$ is a solution to \eqref{eqn:Watson-combined} with no poles or zeros in the physical strip, while $k(\theta)=k(-\theta)=k(2i \pi+\theta)$ captures all the poles and zeros. Additional ``minimality" assumptions regarding the absence of zeros away from threshold, $\theta=0$, are often made \cite{Karowski:1978vz} and can be checked against explicit perturbative equations. However, it is worth noting that this additional assumption selects specific solutions, and hence corresponds to specific operators. 

To determine the ``minimal solution", $f_{\rm min}$, a standard method is by contour integration \cite{Karowski:1978vz} (for a more recent application of the same argument to the $\algSU(N)$ PCM see \cite{Cubero:2012xi}). However, this is equivalent to the formal solution corresponding to an infinite product of S-matrices:
\< \label{eqn:ff-formal-sol}
  f_{\rm min}(\theta)
  =\prod_{n=1}^\infty \Smat(-\theta+2 i n \pi)
  =\prod_{n=1}^\infty \Smat^{-1}(\theta-2 i n \pi)~.
\>
While this product is strictly divergent and so must be interpreted with care, if nothing else, it can be used to find a candidate solution which can then be verified\footnote{The contour method for the minimal form factors is also formal in the sense that it produces divergences. These can be removed by calculating the logarithmic derivative and then upon integration setting the additive constant to be finite.}. We will start by reviewing this strategy for the sine-Gordon model and show that it yields the minimal form factor solution originally due to Weisz \cite{Weisz:1977ii}. Afterwards we will apply it to the string world-sheet theory.

%%%%%%%%%%%%%%%%%%%%%%%%%%%%%%%%%%%%%%%%%%%%%%%%%%%%%%%%%%%%%
\subsection{Sine-Gordon Theory}

\paragraph{Soliton-soliton form factor}

The sine-Gordon soliton-soliton S-matrix is \cite{Zamolodchikov:1977py} 
\<
\Smat_{ss}(\theta)=\prod_{k=1}^\infty \frac{
 \Gamma\bigsbrk{\frac{1}{\lsg}\lrbrk{2k  +\frac{i \theta}{\pi}}} \Gamma\bigsbrk{1+\frac{1}{\lsg}\lrbrk{2k-2+\frac{i \theta}{\pi}}}
 \Gamma\bigsbrk{\frac{1}{\lsg}\lrbrk{2k-1-\frac{i \theta}{\pi}}} \Gamma\bigsbrk{1+\frac{1}{\lsg}\lrbrk{2k-1-\frac{i \theta}{\pi}}}}{
 \Gamma\bigsbrk{\frac{1}{\lsg}\lrbrk{2k  -\frac{i \theta}{\pi}}} \Gamma\bigsbrk{1+\frac{1}{\lsg}\lrbrk{2k-2-\frac{i \theta}{\pi}}}
 \Gamma\bigsbrk{\frac{1}{\lsg}\lrbrk{2k-1+\frac{i \theta}{\pi}}} \Gamma\bigsbrk{1+\frac{1}{\lsg}\lrbrk{2k-1+\frac{i \theta}{\pi}}}}
\>
where $\lsg$ is the coupling constant. 
To find a useful integral form of the S-matrix we can use Malmst\'en's formula 
\<
 \ln \Gamma(z) = \int^\infty_0 \frac{dt}{t} \left( (z-1)-\frac{1-e^{-(z-1)t}}{1-e^{-t}}\right)e^{-t}
\>
to write the logarithm of the S-matrix as 
\<
 \ln \Smat_{ss}(\theta) \eq \sum_{k=1}^\infty \int^\infty_0 \frac{dt}{t} 2e^{-2 k t/\lsg} \frac{(e^t-e^{t/\lsg})(1-e^{t/\lsg})}{e^t-1} \sinh\frac{t\theta}{i \pi\lsg} \nln[1mm]
\eq \int^\infty_0 \frac{dt}{t} \lrbrk{ -1+\coth \frac{t}{2} \tanh \frac{t}{2\lsg} } \sinh\frac{t\theta}{i \pi\lsg} \nln[1mm]
\eq \int^\infty_0 \frac{dt}{t} \, \frac{\sinh\tfrac{(1-\lsg)t}{2} }{ \sinh \tfrac{\lsg t}{2} \cosh \tfrac{t}{2} } \, \sinh\frac{t\theta}{i \pi}~,
\>
where in the last line we rescaled the parameter $t$ by $\lsg$. According to \eqref{eqn:ff-formal-sol}, we can find from this the logarithm of the form factor as the sum
\< \label{eqn:ff-as-div-sum}
 \ln f_{\rm min}(\theta) = \sum_{n=1}^\infty \ln \Smat_{ss}(-\theta+2 in\pi) = \int^\infty_0 \frac{dt}{t} \, h(t) \, \sum_{n=1}^\infty \sinh\Bigbrk{ \frac{t}{i \pi}(-\theta+2in\pi) } ~,
\>
where $h(t) = \frac{\sinh(1-\lsg)t/2 }{ \sinh\lsg t/2 \, \cosh t/2 }$. The sum in \eqref{eqn:ff-as-div-sum} is not well defined, in order to obtain a sensible expression we separate the $\sinh$-function into two exponentials and obtain two convergent series (though convergent for different values of the rapidity). Performing the summations we find
\<
 \ln f_{\rm min}(\theta)
 = -\int^\infty_0 \! \frac{dt}{t} \: h(t) \, \frac{\cosh\bigbrk{(\theta-i\pi)\tfrac{t}{i \pi}}}{2\sinh t}
 = C + \int^\infty_0 \! \frac{dt}{t} \: h(t) \, \frac{\sinh^2\bigbrk{(\theta-i \pi)\tfrac{t}{2i \pi}}}{\sinh t}~,
\>
where $C$ is independent of $\theta$. While the individual steps in the derivation are merely formal manipulations, it is straightforward to check that the final answer is indeed a solution of Watson's equations with the required properties and moreover the second form is exactly the solution of Weisz \cite{Weisz:1977ii} and Karowski and Weisz \cite{Karowski:1978vz}. 

\bigskip
 
We can also write this product solution in a notation similar to Vieira and Volin \cite{Vieira:2010kb}. Defining a shift operator $\shift$ by the action $\shift h(\theta) = h(\theta+2 i \pi)$ and $h(\theta)^{g(\shift)} = \exp(g(\shift) \ln h(\theta))$ for some function $g(\shift)$, we can write \eqref{eqn:Watson-combined} as
\< \label{eqn:Watson-formal}
  f_{\rm min}(\theta)^{\shift-1}=\Smat(-\theta)~.
\>
Thus we can solve formally the equation by writing
\<
f_{\rm min}(\theta) = \Smat(-\theta)^{\tfrac{1}{\shift-1}}
          = \prod_{n=1}^\infty \Smat(-\theta)^{\shift^{-n}}
					= \prod_{n=1}^\infty \Smat(-\theta+2 in \pi)~,
\>
where we have expanded the exponent as if $\shift>1$. We could equally have expanded in $\shift<1$ and found
an alternative expression for the solution
\<
f_{\rm min}(\theta) = \prod_{n=0}^\infty \Smat(-\theta)^{-\shift^{n}}
          = \Smat(\theta) \prod_{n=1}^\infty \Smat(\theta+2 in \pi)~,
\>
which is related to the first expansion by $f(\theta) = \Smat(\theta)f(-\theta)$, see \eqref{eqn:Watson}. 

\paragraph{Breather-Breather form factor}
As we will see below scattering in the near-flat or Malda\-cena-Swanson limit  of the string world-sheet theory is closely related to  breather-breather scattering in sine-Gordon theory and so, as a warm up, it is useful to solve the functional equations in sine-Gordon theory perturbatively. We will consider the sine-Gordon breather-breather S-matrix, which is given by 
\<
\Smat_{bb}=\frac{\sinh\theta+i \sin \pi \nu }{\sinh\theta-i \sin \pi \nu}~,
\>
were $\nu$ is the coupling, which we will take to be small in our perturbative expansion. This S-matrix can be written as an integral \cite{Babujian:1998uw}
\<
\Smat_{bb} = -\exp \int^\infty_0 \frac{dt}{t} \, \frac{2\cosh (\nu-\tfrac{1}{2}) }{\cosh\frac{t}{2}}\sinh\frac{t\theta }{i\pi}~.
\>
Expanding at small coupling, $\nu \rightarrow 0$ we find
\<
\Smat_{bb} = \Smat^{(0)}+\Smat^{(1)}+\dots
           = 1+ 2i\pi \nu \csch\theta+\order(\nu^{2})~.
\>
Correspondingly we can write
\<
\Smat^{(1)} = -  2\nu \int_0^\infty dt~ \tanh\frac{t}{2}\, \sinh\frac{t \theta}{i\pi}~.
\>
We can explicity perform the integral by contour integration. There are poles at $t=i \pi (2n+1)$ for $n\in\Integers$. We extend the integration to the entire real line and split the $\sinh$ into two factors $e^{i t \theta/\pi}$ and $e^{-i t \theta/\pi}$. For $\Re\theta>0$, we can close the contour for the first term in the upper half plane picking up the poles at $n=0,1,2,\dots$ with residues
\<
\nu e^{-(2n+1)\theta}~,
\>
while for the second term, we close the contour in the lower half plane picking up the poles at $n=-1, -2, \dots$ with residues $-\nu e^{(2n+1)\theta}$. Taking into account the different orientation of the contours and summing over all poles we find 
\<
\Smat^{(1)}= 2\pi i \nu \csch \theta~,
\>
as expected. For the perturbative form factor 
\<
f_{\mathrm{min}}(\theta)=f^{(0)}_{\mathrm{min}}(\theta)+f^{(1)}_{\mathrm{min}}(\theta)+\dots
\>
the formula \eqref{eqn:ff-formal-sol} reduces at first order to
\<
f^{(1)}_{\mathrm{min}}(\theta)
=\sum_{n=1}^\infty \Smat^{(1)}(-\theta+2 i\pi n)
=\frac{\nu }{2}\int_0^\infty \!dt\: \frac{\cosh\bigbrk{(\theta - i\pi)\tfrac{t}{i\pi}}}{\cosh^2 \tfrac{t}{2}}~.
\>
This integral can again be done by contour integration; there are poles at $t=(2n+1)i \pi$. The final answer is 
\<
f^{(1)}_{\mathrm{min}}(\theta)=- \nu (\theta - i\pi) \csch\theta~.
\>
Now, let us turn to the near-flat limit of the world-sheet theory.

%%%%%%%%%%%%%%%%%%%%%%%%%%%%%%%%%%%%%%%%%%%%%%%%%%%%%%%%%%%%%%%%%%%%%%%%%%%
%%%%%%%%%%%%%%%%%%%%%%%%%%%%%%%%%%%%%%%%%%%%%%%%%%%%%%%%%%%%%%%%%%%%%%%%%%%
\subsection{Near-Flat World-Sheet Theory}

The world-sheet theory in light-cone gauge is not Lorentz invariant and so the previous methods are not directly applicable. However,  they can be generalised and in particular the axioms be formulated straightforwardly \cite{Klose:2012ju}. In the case of the Maldacena-Swanson limit the world-sheet theory becomes ``almost" Lorentz invariant. In terms of the rapidity variables, $\theta_i$, defined by $\eta_i=e^{\theta_i}$, we can write the near-flat S-matrix in terms of a rapidity-dependent coupling, 
\<
  \gt = \gamma e^{\theta_1+\theta_2}~,
\>
where $\gamma$ is the loop counting parameter, and the rapidity difference $\theta=\theta_2-\theta_1$. To two-loop order, the S-matrix in the $\algSU(2)$ sector is 
\<
\Smat(\gt, \theta) = 1+2 i \gt \coth \frac{\theta}{2}-2 \gt^2 \coth^2 \frac{\theta}{2}+\frac{4 i \gt^2}{\pi}(1-\theta \coth\theta)\csch\theta + \order(\gt^3)~.
\>
This can be written, again to two loops, as
\<
\ln \Smat(\gt, \theta) = 2 i \gt \coth \frac{\theta}{2}+\frac{4 i \gt^2}{\pi}(1-\theta \coth\theta)\csch\theta + \order(\gt^3)~,
\>
which can be written in a convenient integral form. Before proceeding to that step however, it is interesting to note that considering just the BDS \cite{Beisert:2004hm} part of the S-matrix, in the near-flat limit this becomes, to all orders in $\gt$, 
\<
\Smat_{\rm BDS} = 1-\frac{2}{1-\ihalf \bigbrk{\gt + \frac{1}{\gt}} \sinh \theta } 
            = 1-\frac{2}{1-\frac{i}{\beta}\sinh \theta}
						= \frac{\sinh\theta -i \beta}{\sinh \theta + i \beta}~,
\>
where we have written the rapidity-dependent coupling as $\beta = \frac{2\gt}{1+\gt^2}$. It is interesting to observe the similarity of this S-matrix to that for breathers in sine-Gordon theory 
\<
\Smat_{bb} = \frac{\sinh\theta +i \sin \pi \nu}{\sinh \theta -i \sin \pi \nu}
       = -\exp\int_0^\infty \frac{dt}{t}\frac{2 \cosh (1/2-\nu)t}{\cosh t/2}\sinh\frac{t\theta}{i \pi}~.
\>
Returning to the full S-matrix, but only to two loops, we can write the logarithm as
\<
\ln \Smat(\gt, \theta) = -\frac{4 \gt}{\pi}\int_0^\infty \!dt\: \coth t \sinh \frac{t\theta}{i \pi}
- \frac{2 \gt^2}{\pi^2} \int_0^\infty \!dt\: t\, \tanh^2\frac{t}{2}\sinh \frac{t\theta}{i \pi}+{\cal O}(\gt^3)~.
\>
In this formula we have left the S-matrix invariant under shifts of the effective coupling, $\ln \tilde \gamma \rightarrow \ln \tilde \gamma+2\pi i $, i.e. in the sum of rapidities while extending it beyond the physical region as a function of  the rapidity difference. This formula is now  of the same form as the Lorentz invariant sine-Gordon case and with this motivation we will apply the above methods. 

Once again using the relation 
\<
\exp \ln f_{\mathrm{min}}(\theta)=\exp\left(\sum_{n=1}^\infty \ln \Smat(\gt, -\theta+2 in \pi)\right)
\>
we find that
\<
f_{\mathrm{min}}(\theta) \eq 1
- \frac{\gt}{\pi} (\theta - i\pi) \coth\frac{\theta}{2}
+ \frac{\gt^2}{2\pi^2}(\theta - i\pi)^2 \coth^2\frac{\theta}{2}\nl
- \frac{\gt^2}{\pi^2}(\theta - i\pi)\bigbrk{2 -(\theta-i\pi) \coth \theta} \csch\theta + \order(\gt^3)~.
\>
This can be compared with the two-loop perturbative calculation \eqref{eqn:nfs-tree-and-one} and \eqref{eqn:nfs-two} (rewritten in terms
of $\gt$ and $\theta$)
\<
f_{\rm pert}(\eta_1,\eta_2)&=&\frac{1}{2\sqrt{\eta_1\eta_2}}\Big[1
- \frac{\gt}{\pi} (\theta - i\pi) \coth\frac{\theta}{2}
+ \frac{\gt^2}{2\pi^2}(\theta - i\pi)^2 \coth^2\frac{\theta}{2}\nn\\
& & ~~~~~~~~~~~ - \frac{\gt^2}{\pi^2}(\theta - i\pi)\bigbrk{2 -(\theta-i\pi) \coth \theta} \csch\theta\nn\\
&&~~~~~~~~~~~ -\gt^2\left(\frac{1}{6}\cosh\theta+\frac{1}{2}\frac{1}{1-\sech \theta}+\left(\frac{1}{\pi}-\frac{1}{12}\right)\cosh 2\theta\right)\Big]
\>
and it can be seen that the minimal solution correctly reproduces all the terms involving bubble integrals, or correspondingly, the logarithmic terms, here appearing as $(\theta - i\pi)$. In fact, we can write the perturbative expression as
 \<
f_{\rm pert}(\eta_1,\eta_2)&=&\frac{1}{2\sqrt{\eta_1\eta_2}}k(\gt, \theta)f_{\mathrm{min}}(\gt, \theta)
\>
with 
\<
k(\gt, \theta)=1-\gt^2\left(\frac{1}{6}\cosh\theta+\frac{1}{2}\frac{1}{1-\sech \theta}+\left(\frac{1}{\pi}-\frac{1}{12}\right)\cosh 2\theta\right)
\>
which is indeed even and periodic in $\theta$. However, it is does not appear to follow from an obvious ``minimality" condition such as used in relativistic theories. It is possible that we need to correct the operator $Y^2$ at higher orders and that such a correctly defined operator would satisfy minimality. In either case, it would certainly be interesting to better understand any constraints, such as those following from bound state singularities,  that would allow one to determine this function without recourse to perturbation theory. 

%%%%%%%%%%%%%%%%%%%%%%%%%%%%%%%%%%%%%%%%%%%%%%%%%%%%%%%%%%%%%%%%%%%%%%%%%%%
%%%%%%%%%%%%%%%%%%%%%%%%%%%%%%%%%%%%%%%%%%%%%%%%%%%%%%%%%%%%%%%%%%%%%%%%%%%
\subsection{Near-Plane-Wave World-Sheet Theory}

It is also interesting to consider the perturbative form factors in the near-BMN or near-plane wave limit discussed in \secref{sec:npw-pert}. We again focus on a single $\algSU(2)$ sector. The world-sheet 
tree-level S-matrix was perturbatively calculated for a class of light-cone type gauges  in \cite{Klose:2006zd}, and for the scattering $YY\rightarrow YY$ one finds
\<
\label{eq:su2_tree_smatrix}
\Smat= 1 + \frac{i\pi}{\sqrt{\lambda}}\left( \frac{(p_1+p_2)^2}{\epsilon_2 p_1-\epsilon_1 p_2} + (1-2a)(\epsilon_2 p_1-\epsilon_1 p_2) \right)~,
\>
where $a$ characterizes the gauge-fixing. Now, we work with the usual rapidity parameters \eqref{eqn:usual-rapitidies} and the combinations $\theta=\theta_2-\theta_1$ and $\tilde\theta=\theta_2+\theta_1$. As we will see, at least to one-loop and for the $a=\tfrac{1}{2}$ gauge, the sum of rapidities can be combined with the coupling such that we can write the S-matrix in an integral form much as in sine-Gordon and almost exactly parallel to the near-flat case. Then the same trick for finding a solution to Watson's equations can be employed and we find a one-loop minimal form factor that matches the perturbative Feynman diagram calculation.

In terms of $\theta$ and $\tilde\theta$ the S-matrix, \eqref{eq:su2_tree_smatrix}, is 
\<
  \Smat=1 - \frac{i \pi }{\sqrt{\lambda}}\lrbrk{ 2 \coth\frac{\theta}{2} \sinh^2 \frac{\tilde \theta}{2} + (1-2a) \sinh\theta }
\>
so that with $a=\tfrac{1}{2}$ we can write (to leading order)
\<
  \ln \Smat = \frac{4}{\sqrt{\lambda}} \sinh^2 \frac{\tilde \theta}{2} \int_0^\infty\!dt\: \coth t\; \sinh\frac{t \theta}{i \pi} \; .
\>
Thus, valid to order $1/\sqrt{\lambda}$, we find for the minimal solution 
\<
  f_{\rm min} = \exp \lrbrk{ -\frac{2}{\sqrt{\lambda}} \sinh^2 \frac{\tilde \theta}{2} \int_0^\infty\!dt\: \coth t \, \; \frac{\cosh\bigbrk{(\theta - i \pi)\frac{t}{i\pi}}}{\sinh t} }
\>
or
\< \label{eqn:ff-npw-Watson}
  f_{\rm min} = 1 + \frac{1}{\sqrt{\lambda}} (\theta - i\pi) \coth\frac{\theta}{2} \sinh^2\frac{\tilde{\theta}}{2} + \order(\lambda^{-1}) \; .
\>
which (up to the overall wave-function factor) agrees with the Feynman diagram computation, \eqref{eqn:ff-ren-rapid-a12}, in the $a=1/2$ gauge.

%%%%%%%%%%%%%%%%%%%%%%%%%%%%%%%%%%%%%%%%%%%%%%%%%%%%%%%%%%%%%%%%%%%%%%%%%%%
%%%%%%%%%%%%%%%%%%%%%%%%%%%%%%%%%%%%%%%%%%%%%%%%%%%%%%%%%%%%%%%%%%%%%%%%%%%
\section{Gauge Theory Structure Constants from Form Factors}
\label{sec:gauge-theory}

We wish to compare the world-sheet form factors calculated above with gauge theory structure constants. The motivation for this identification comes from the fact that for specific gauge theory operators the tree-level structure constants can be related to spin-chain matrix elements of specific operators, see e.g. \cite{Roiban:2004va}. The OPE coefficients for three operators, $O^a$, $O^b$, and $O^c$, naturally have the expansion at small 't Hooft coupling
\<
C^{abc}=c_0^{abc}(1+\lambda c_1^{abc}+\dots)~,
\>
where the leading term $c_0^{abc}$ is given by free field contractions. This leading term can thus be related to a spin-chain matrix element where two of the gauge theory operators serve as in- and out-states, say $O^a \to \bra{a}$ and $O^c \to \ket{c}$, and the third as a spin-chain operator $O^b \to \mathcal{O}^b$. The example which is relevant to our considerations is the $\algSU(2)$ sector comprising the complex scalars $Z$ and $Y$, which is described at one-loop by the spin-$1/2$ XXX spin-chain. The vacuum state is naturally identified with the normalised BPS-state, 
\<
\frac{1}{\sqrt{L_c}}\tr(Z^{\Lc}) \leftrightarrow \ket{0}_{\Lc}= \ket{\uparrow\uparrow\dots\uparrow}~,
\>
where $\Lc$ denotes the spin-chain length. For operators with equal numbers of holomorphic and antiholomorphic fields an explicit representation in terms of the usual spin-chain operators can be found, e.g.
\<
O=\tr(ZZ\bar Y\bar Y)\leftrightarrow {\cal O}=\sum_{j=1}^{\Lc} S_{+,j}S_{+,j+1}~,
\>
where $S_{+,j}$ is the spin raising operator acting on site $j$. Famously, the XXX spin-chain can be solved by the Bethe Ansatz (see e.g. \cite{Faddeev:1996iy} for a review) and in general the inverse scattering method expresses local spin-chain operators in terms of the transfer matrix. However, we will not need the full power of this method for our considerations as will consider states with at most two excitations: $\ket{\psi(p_1,p_2)}_{\Lc}$. Such states  are eigenvectors of the transfer matrix when the momenta satisfy the Bethe equations (BE) and after further imposing the condition of vanishing total momentum, $p_1=-p_2$, the spin-chain state corresponds to the BMN single trace operators of the gauge theory \cite{Beisert:2002tn}
\<
O=\frac{1}{\sqrt{\Lc-1}}\sum_{j=0}^{\Lc-2}\cos\frac{\pi n(2j+1)}{\Lc-1}\tr(YZ^iYZ^{\Lc-2-j})~,
\>
where $n$ is the mode number that characterises the solution of the BE. 
We will focus on the case where one operator is the vacuum state $O^a\sim \tr(\bar Z^{L_c})$, a second is the short operator $O^b=\tr(ZZ\bar Y\bar Y)$ and the third is a BMN operator with two-impurities $O^c\sim {\rm}(Y^2Z^{L_c-2})$. To this end we calculate the spin-chain
matrix element
\<
\label{eq:struct_const_spin_chain}
c_0^{abc}={}_{\Lc}\bra{0} S_{+,j} S_{+,j+1}\ket{\psi(p_1,p_2)}_{\Lc}
\>
where the excitation momenta satisfy the BE (but for generality we will not impose vanishing total momentum).

Our aim is compare the thermodynamic expansion of this matrix element with the two-particle world-sheet form factor. The thermodynamic
limit of the spin-chain is described by a Landau-Lifshitz model (LL) \cite{Fradkin:1991nr, Affleck:1988nt}. It is well known that the world-sheet action, including loop-effects, can be matched to the LL action by appropriate field redefinitions \cite{Kruczenski:2003gt, Kruczenski:2004kw} (we review this matching in \appref{app:LL}) and thus it is unsurprising that agreement is found in the appropriate limits. However, it is useful to consider the explicit matching as it highlights several important feature of the gauge theory/string theory comparison that will be relevant more generally. The key issue is that while the spin-chain matrix elements correspond to periodic, finite length states the world-sheet form factors are calculated in infinite volume. Thus, the first step is to recompactify the string world-sheet so that it has finite length, say $\Ls$. We propose that finite-volume effects to all polynomial orders in $1/\Lc$ can then be accounted for by: 
\begin{itemize}
\item[(i)]
demanding that the momenta satisfy the string Bethe equations 
\item[(ii)]
properly taking into 
account the normalization of the states. 
\end{itemize}
This proposal is quite similar to the analogous procedure used in relativistic integrable models as 
described in \cite{Pozsgay:2007kn, Pozsgay:2008bf, Pozsgay:2013jua}. Naturally there could also be exponential corrections which will require more involved techniques such as the TBA but we leave such considerations to the future.

%%%%%%%%%%%%%%%%%%%%%%%%%%%%%%%%%%%%%%%%%%%%%%%%%%%%%%%%%%%%%%%%%%%%%%%%%%%
%%%%%%%%%%%%%%%%%%%%%%%%%%%%%%%%%%%%%%%%%%%%%%%%%%%%%%%%%%%%%%%%%%%%%%%%%%%
\subsection{Spin-Chain Form Factors}

In order to check our proposal in at least one concrete case, we review some expressions regarding the spin-chain form factor
\< \label{eqn:spinchain-ff-def}
  f_{\rm spin}(p_1,p_2) = \bra{0} S_{+,1} S_{+,2} \ket{\hat{\psi}(p_1,p_2)}
\>
for the two-particle state $\ket{\hat{\psi}(p_1,p_2)}$, which is the normalized version of
\< \label{eqn:spin-ket-not-normalized}
  \ket{\psi(p_1,p_2)} = \sum_{1\le x_1<x_2\le \Lc} \psi(p_1,p_2)_{x_1,x_2} \ket{x_1,x_2} ~.
\>
We take the wave-function to be
\< \label{eqn:two-spin-wave-function}
  \psi(p_1,p_2)_{x_1,x_2} = e^{ip_1x_1 + ip_2x_2 + \ihalf\Theta_{12}} + e^{ip_2x_1 + ip_1x_2 - \ihalf\Theta_{12}} ~,
\>
where the phase-shift $\Theta_{12} \equiv \Theta_{\mathrm{c}}(p_1,p_2)$ is given in terms of the Heisenberg S-matrix by
\< \label{eqn:phase-shift}
  e^{i\Theta_{\mathrm{c}}(p_1,p_2)} = \Smat(p_1,p_2) = - \frac{e^{i(p_1+p_2)} - 2 e^{ip_1} + 1}{e^{i(p_1+p_2)} - 2 e^{ip_2} + 1} ~.
\>
In order to normalize the state, we compute the norm $\mathcal{N}_{\mathrm{c}}(p_1,p_2) = \braket{\psi(p_1,p_2)}{\psi(p_1,p_2)}$ and then divide the state by $\sqrt{\mathcal{N}_{\mathrm{c}}}$. This does \emph{not} fix the state completely, but only up to an overall phase which is arbitrary and as we will see this contribution will not match the corresponding term in the world-sheet form factor. The form factor is then given by
\< \label{eqn:ff-spin-normalized}
  f_{\rm spin}(p_1,p_2) = \frac{\psi(p_1,p_2)_{1,2}}{\sqrt{\mathcal{N}_{\mathrm{c}}(p_1,p_2)}} ~.
\>

%%%%%%%%%%%%%%%%%%%%%%%%%%%%%%%%%%%%%%%%%%%%%%%%%%%%%%%%%%%%%%%%%%%%%%%%%%%
%%%%%%%%%%%%%%%%%%%%%%%%%%%%%%%%%%%%%%%%%%%%%%%%%%%%%%%%%%%%%%%%%%%%%%%%%%%
\paragraph{Mode numbers}
As mentioned, spin-chain states corresponding to eigenoperators of the dilatation generator have momenta satisfying the BE, i.e. 
\<
  p_1 \Lc = \Theta_{\mathrm{c}}(p_1,p_2) + 2\pi n_1
	\comma 
  p_2 \Lc = -\Theta_{\mathrm{c}}(p_1,p_2) + 2\pi n_2
	~.
\>
In making the comparison with the world-sheet theory, we are most interested in the
thermodynamic limit and in a large-$\Lc$ expansion the momenta are given by
\< \label{eqn:spin-chain-momenta-modes}
  p_1 = \frac{2\pi n_1}{\Lc} - \frac{4\pi}{\Lc^2} \frac{n_1 n_2}{n_1-n_2} + \order(\Lc^{-3})
  \comma
  p_2 = \frac{2\pi n_2}{\Lc} + \frac{4\pi}{\Lc^2} \frac{n_1 n_2}{n_1-n_2} + \order(\Lc^{-3})
	~.
\>

%%%%%%%%%%%%%%%%%%%%%%%%%%%%%%%%%%%%%%%%%%%%%%%%%%%%%%%%%%%%%%%%%%%%%%%%%%%
%%%%%%%%%%%%%%%%%%%%%%%%%%%%%%%%%%%%%%%%%%%%%%%%%%%%%%%%%%%%%%%%%%%%%%%%%%%
\paragraph{Normalization}
We can compute the norm directly by performing the sum 
\<
  \braket{\psi(p_1,p_2)}{\psi(p_1,p_2)} \eq \Lc(\Lc-1) + e^{-i\Delta p(\Lc-1)+i\theta_{12}} \sum_{x=1}^{\Lc-1} x e^{i\Delta p(x-1)} \nl
	\hspace{15mm} + e^{i\Delta p(\Lc-1)-i\theta_{12}} \sum_{x=1}^{\Lc-1} x e^{-i\Delta p(x-1)}
\>
where $\Delta p = p_1-p_2$ and, using the Bethe equations to replace the factors of the form $e^{ip_i\Lc}$, we find 
\< \label{eqn:sc-scalar-product-exact}
  \braket{\psi(p_1,p_2)}{\psi(p_1,p_2)} =
	\Lc \lrbrk{ \Lc - \frac{2 (2 - \cos p_1 - \cos p_2)}{3 - 2 ( \cos p_1 + \cos p_2 ) + \cos(p_1+p_2)} }~.
\>
As is well known, the same result can also be found from a determinant expression \cite{Gaudin:1971zza, Gaudin:1981norm} (see \appref{sec:sc-norm-from-Jacobian}).

%%%%%%%%%%%%%%%%%%%%%%%%%%%%%%%%%%%%%%%%%%%%%%%%%%%%%%%%%%%%%%%%%%%%%%%%%%%
%%%%%%%%%%%%%%%%%%%%%%%%%%%%%%%%%%%%%%%%%%%%%%%%%%%%%%%%%%%%%%%%%%%%%%%%%%%
\paragraph{Final result}
Using the expansion \eqref{eqn:spin-chain-momenta-modes}, the wave-function \eqref{eqn:two-spin-wave-function} becomes
\<
  \psi(p_1,p_2)_{1,2} = 2 + \frac{6i\pi}{\Lc} (n_1+n_2) + \order(\Lc^{-2}) ~,
\>
and the scalar product \eqref{eqn:sc-scalar-product-exact} is
\<
  \braket{\psi(p_1,p_2)}{\psi(p_1,p_2)} =	\Lc^2 - 2\Lc \frac{n_1^2+n_2^2}{(n_1-n_2)^2} + \order(\Lc^0)~,
\>
leading to the normalization factor
\< \label{eqn:sc-norm-factor}
  \mathcal{N}_{\mathrm{c}}^{-1/2} = \frac{1}{\Lc} + \frac{1}{\Lc^2} \frac{n_1^2 + n_2^2}{(n_1-n_2)^2} + \order(\Lc^{-3}) ~.
\>
Combining these, the final expression for the spin-chain form factor, \eqref{eqn:ff-spin-normalized}, to subleading order in $1/\Lc$ is
\<
\label{eq:spin_chain_ff}
  f_{\rm spin}(p_1,p_2) = \frac{2}{\Lc} + \frac{6i\pi}{\Lc^2} (n_1+n_2)  + \frac{2}{\Lc^2} \frac{n_1^2 + n_2^2}{(n_1-n_2)^2} + \order(\Lc^{-3}) ~.
\>
As discussed above, this expression is related to the tree-level structure constants \eqref{eq:struct_const_spin_chain} and
it is this expression we wish to relate to the string theory calculation.

%%%%%%%%%%%%%%%%%%%%%%%%%%%%%%%%%%%%%%%%%%%%%%%%%%%%%%%%%%%%%%%%%%%%%%%%%%%
%%%%%%%%%%%%%%%%%%%%%%%%%%%%%%%%%%%%%%%%%%%%%%%%%%%%%%%%%%%%%%%%%%%%%%%%%%%
\subsection{World-Sheet Form Factors in Finite Volume}

For the form factor calculation, and following the procedure outlined above, we must recompactify the string world-sheet. Prior to decompactification, for the general $a$-gauge light-cone string action, the world-sheet length, $\Ls$, is related to the string energy, $E$, and angular momentum on the
S$^5$,  $J$, via the relation
\< \label{eqn:world-sheet-length}
 \Ls = 2\pi \lrsbrk{ (1-a)\frac{J}{\sqrt{\lambda}} + a\frac{E}{\sqrt{\lambda}} } ~.
\> 
Replacing $E$ in \eqref{eqn:world-sheet-length} by the expression for the light-cone energy
\<
  E = J + \sum_i \sqrt{1 + p_i^2}~.
\>
one obtains
\<
  \Ls = \frac{2\pi}{\sqrt{\lambda}} \lrsbrk{ J  + a \sqrt{1+p_1^2} + a \sqrt{1+p_2^2} }
\>
in the case of two momenta. In principle, it is important to note that the momenta themselves depend on the world-sheet length, see \eqref{eqn:string-momenta-modes-Ls}. However, being interested only in the limit of large $J$ and large $J/\sqrt{\lambda}$, it is sufficient to approximate $\sqrt{1+p_i^2}$ by $1$. Thus we are left with
\< \label{eqn:length-conversion}
  \Ls = \frac{2\pi}{\sqrt{\lambda}} \bigsbrk{ J + 2a + \order(J^{-1})} = \frac{2\pi}{\sqrt{\lambda}} \bigsbrk{ \Lc - 2(1-a) + \order(\Lc^{-1}) } ~.
\>
If we wished to express our final answers in terms of the gauge theory R-charge which is dual to the string angular momentum, $J$, it would be natural to work in $a=0$ gauge. Alternatively, to match with the Landau-Lifshitz theory describing the thermodynamic limit of the spin-chain we need to express our answers in terms of the spin-chain length, $\Lc$, which is given by the R-charge, $J$, plus the number of excitations, $M = 2$, we work in $a=1$ gauge. Further, as can be seen from the match with the LL-action, \appref{app:LL}, to get the correct identifications we rescale the fields and world-sheet coordinates so that the length is $\Ls=\tfrac{2\pi {\Lc}}{ \sqrt{\lambda}}$ while the loop counting parameter will effectively be  $1/ {\Lc}$. 

\paragraph{Mode numbers} Next we need 
to express the excitation momenta in terms of the mode numbers characterising solutions to the string Bethe equations for a 
world-sheet of length $\Ls$: 
\<
  p_1 \Ls = \Theta_{\mathrm{s}}(p_1,p_2) + 2\pi n_1
	\comma 
  p_2 \Ls = -\Theta_{\mathrm{s}}(p_1,p_2) + 2\pi n_2
\>
where the worldsheet S-matrix\footnote{The S-matrix appearing in the Bethe ansatz is related to that calculated directly from the worldsheet theory by $ \Smat={\cal P}_g P^u_{p_1p_2}S^B$, where ${\cal P}_g$ is the graded permutation operator and $P^u_{p_1p_2}$ exchanges the excitation momenta.} in the relevant $\algSU(2)$ sector is given by $S^B(p_1,p_2)=e^{i\Theta_{\mathrm{s}}(p_1,p_2)}$ with the phase shift to leading order in $1/\sqrt{\lambda}$ is given by
\<
  \Theta_{\mathrm{s}}(p_1,p_2) = - \frac{\pi}{\sqrt{\lambda}} \left( \frac{(p_1+p_2)^2}{\epsilon_2 p_1-\epsilon_1 p_2} + (1-2a)(\epsilon_2 p_1-\epsilon_1 p_2) \right) + \order(\lambda^{-1})~.
\>
The solution for the momenta to order $1/\sqrt{\lambda}$ is
\< \label{eqn:string-momenta-modes-Ls}
  p_1 = \frac{2 \pi n_1}{\Ls} - \frac{4\pi^2}{\sqrt{\lambda}\Ls^2} \frac{n_1^2 + n_2^2 - a(n_1-n_2)^2}{n_1 - n_2} + \order(\lambda^{-1})~,
\>
and similar for $p_2$. Inserting these expressions into the formulas \eqref{eqn:ff-npw-tree} and \eqref{eqn:ff-ren-rapid} for the form factor one obtains to order $1/\sqrt{\lambda}$, 
\< \label{eqn:ff-npw-Ls}
  f^{(0)}(n_1,n_2) \eq \frac{1}{2} - \frac{\pi^2}{2\Ls^2} (n_1^2+n_2^2) + \order(\Ls^{-3}) ~, \\ 
  f^{(1)}_{\mathrm{ren}}(n_1,n_2) \eq \frac{i\pi^2}{2\sqrt{\lambda}\Ls} \lrsbrk{ \frac{(n_1+n_2)^2}{n_1-n_2} + (1-2a) (n_1-n_2) } \nl
	+ \frac{2\pi^2}{\sqrt{\lambda}\Ls^2} \biggsbrk{ (n_1^2+n_2^2) - a (n_1-n_2)^2 } + \order(\Ls^{-3}) ~. \nn
\>
We will in fact only need to keep terms to order $1/\Ls$ as the subleading terms  correspond to one-loop and higher corrections in the gauge theory expansion. It would certainly be interesting to understand how to match these subleading terms, however this will be left to future work and we will content ourselves with the tree-level gauge theory structure constants. 

%%%%%%%%%%%%%%%%%%%%%%%%%%%%%%%%%%%%%%%%%%%%%%%%%%%%%%%%%%%%%%%%%%%%%%%%%%%
%%%%%%%%%%%%%%%%%%%%%%%%%%%%%%%%%%%%%%%%%%%%%%%%%%%%%%%%%%%%%%%%%%%%%%%%%%%
\paragraph{Normalization}

The second step in relating the form factors to finite volume matrix elements is to include an appropriate density of states factor or, equivalently, to use appropriately normalized states. To motivate this factor we consider the fact that the external two-particle states in the world-sheet theory, $\ket{Y(p_1)Y(p_2)}$, satisfy
\< \label{eqn:normalization-string}
  \braket{Y(p_3)Y(p_4)}{Y(p_1)Y(p_2)} =(2\pi)^2\big[ \delta(p_1-p_3)\delta(p_2-p_4) + \text{crossed channel}\big], 
\>
while the two-magnon states of the spin-chain, $\ket{\hat{\psi}(p_1,p_2)}$, are normalized such that
\< \label{eqn:normalization-spin}
  \braket{\hat{\psi}(p_3,p_4)}{\hat{\psi}(p_1,p_2)} = \delta_{p_1,p_3}\delta_{p_2,p_4} + \text{crossed channel}.
\>
These two ways of normalizing the states are not immediately comparable. Instead, the states should be normalized such that the right hand sides are delta-functions of the mode numbers, i.e. $\delta(n_1-n_3)\delta(n_2-n_4)$ and $\delta_{n_1,n_3}\delta_{n_2,n_4}$, respectively. For the Kronecker-delta function this is trival and is given by $\delta_{p_1,p_3} = \delta_{n_1,n_3}$. However, for the Dirac-delta function the change of variables generates a Jacobian:
\< \label{eqn:delta-conversion}
  \delta(p_1-p_3)\delta(p_2-p_4) = \lrabs{\frac{\partial( p_1, p_2 )}{\partial(n_1,n_2)}}^{-1} \delta(n_1-n_3)\delta(n_2-n_4) \; .
\>
The partial derivatives of the momenta can be computed from \eqref{eqn:string-momenta-modes-Ls} and are given by
\begin{align}
  \frac{\partial p_1}{\partial n_1} &= \frac{2\pi}{\Ls} + \frac{4\pi^2}{\sqrt{\lambda}\Ls^2} \lrsbrk{ \frac{-n_1^2+2n_1n_2+n_2^2}{(n_1-n_2)^2} + a } \,, &
  \frac{\partial p_1}{\partial n_2} &= - \frac{4\pi^2}{\sqrt{\lambda}\Ls^2} \lrsbrk{ \frac{n_1^2+2n_1n_2-n_2^2}{(n_1-n_2)^2} + a } \,, \\
  \frac{\partial p_2}{\partial n_1} &= - \frac{4\pi^2}{\sqrt{\lambda}\Ls^2} \lrsbrk{ \frac{-n_1^2+2n_1n_2+n_2^2}{(n_1-n_2)^2} + a } \,, &
  \frac{\partial p_2}{\partial n_2} &= \frac{2\pi}{\Ls} + \frac{4\pi^2}{\sqrt{\lambda}\Ls^2} \lrsbrk{ \frac{n_1^2+2n_1n_2-n_2^2}{(n_1-n_2)^2} + a } \,, \nn
\end{align}
all up to order $\Ls^{-3}$. The Jacobian (times $(2\pi)^2$ from \eqref{eqn:normalization-string}) then becomes
\<
(2\pi)^2  \lrabs{\frac{\partial( p_1, p_2 )}{\partial(n_1,n_2)}}^{-1} \eq \Ls^2 - \frac{4\pi \Ls}{\sqrt{\lambda}} \lrsbrk{ \frac{2n_1n_2}{(n_1-n_2)^2} + a } + \order(\Ls^0) ~.
\>
This expression  gives the additional normalization factor, ${\cal N}_{\rm s}$, that must be included to interpret the form factor as a finite volume matrix element. The form is naturally reminiscent of the Gaudin expression for the norm of Bethe states in non-relativisitic integrable models \cite{Gaudin:1971zza, Gaudin:1981norm, Korepin:1982gg, 1993qism.book.korepin} and is closely related to the density of states factor found in the relativistic case by Poszgay \cite{Pozsgay:2007kn, Pozsgay:2008bf, Pozsgay:2013jua}.
Including the normalization factor we find to order $1/\sqrt{\lambda}$ and $1/\Ls^2$, 
\<
\label{eq:finite_vol_ff}
\hat{f}(n_1,n_2)&=&\frac{1}{\sqrt{{\cal N}_{\mathrm{s}}} }\Big[f^{(0)}(n_1,n_2)+f^{(1)}(n_1,n_2)\Big]
\nn\\
&=&\frac{1}{2 \Ls}+\frac{\pi}{\sqrt{\lambda}\Ls^2}\left(\frac{2n_1n_2}{(n_1-n_2)^2}+a\right)
+\frac{i\pi^2}{2\sqrt{\lambda}\Ls^2}\left(\frac{(n_1+n_2)^2}{n_1-n_2}+(1-2a)(n_1-n_2)\right)\nn\\
& & +{\cal O}\left(\Ls^{-3}\right)~.
\>

%%%%%%%%%%%%%%%%%%%%%%%%%%%%%%%%%%%%%%%%%%%%%%%%%%%%%%%%%%%%%%%%%%%%%%%%%%%
%%%%%%%%%%%%%%%%%%%%%%%%%%%%%%%%%%%%%%%%%%%%%%%%%%%%%%%%%%%%%%%%%%%%%%%%%%%
\subsection{World-Sheet Form Factors in Spin-Chain variables}

While the expression \eqref{eq:finite_vol_ff} gives the two-particle form factor in finite volume, to make a comparison with the results on the spin-chain i.e. the tree-level gauge theory result, ta number of additional issues must be addressed. Firstly, we must express the answer in terms of spin-chain variables, that is we must use the spin-chain length $\Lc$ rather than the world-sheet length $\Ls$, by using \eqref{eqn:length-conversion}. Expressing the world-sheet momenta \eqref{eqn:string-momenta-modes-Ls} in terms of the spin-chain length mixes the two terms in \eqref{eqn:string-momenta-modes-Ls} and yields
\<
  p_1 = \sqrt{\lambda} \lrsbrk{ 
	      \frac{n_1}{\Lc}
				- \frac{1}{\Lc^2} \lrbrk{ \frac{2n_1n_2}{n_1-n_2} - (1-a)(n_1+n_2) }
				 + \order(\Lc^{-3}) }~.
\>
We note that the dependence on the gauge parameter $a$ drops out, when level-matching is imposed, i.e.\ for $n_1 + n_2 = 0$.  If we instead set $a=1$, then
\< \label{eqn:string-momenta-modes-Lc}
  p_1 = \sqrt{\lambda} \lrsbrk{ \frac{n_1}{\Lc} - \frac{1}{\Lc^2} \frac{2 n_1 n_2}{n_1 - n_2}  + \order(\Lc^{-3}) }~,
\>
which equals the spin-chain momentum, \eqref{eqn:spin-chain-momenta-modes}, up to an overall factor of $\frac{2\pi}{\sqrt{\lambda}}$. 
This is a second issue which corresponds to the fact that in order to compare dimensionful quantities, such as the normalised form factors,  between the world-sheet theory 
and the spin-chain we must rescale by such a factor. 

The form factor normalization in spin-chain variables is
\<
{\cal N}_{\mathrm{s}}=(2\pi)^2  \lrabs{\frac{\partial( p_1, p_2 )}{\partial(n_1,n_2)}}^{-1} 
	\eq \frac{(2\pi)^2L^2_c}{\lambda}\lrsbrk{ 1 - \frac{2}{\Lc}  \left( \frac{2(n_1^2-n_1n_2+n_2^2)}{(n_1-n_2)^2} - a \right) + \order(\Lc^0)}~.
\>
We note that the the normalisation is dimensionful from the world-sheet perspective and so before comparing to the spin-chain
one should perform a rescaling by $\frac{\lambda}{(2\pi)^2}$~.
Hence, we should we should multiply the form factor \eqref{eqn:ff-npw-Ls} by the extra  factor
\< \label{eqn:npw-norm-factor}
\frac{2\pi}{\sqrt{\lambda}}\frac{1}{\sqrt{\mathcal{N}_{\mathrm{s}}}}
  \eq \frac{1}{\sqrt{\lambda}}  \lrabs{\frac{\partial( p_1, p_2 )}{\partial(n_1,n_2)}}^{1/2} 
	=   \frac{1}{\Lc} + \frac{1}{\Lc^2} \lrsbrk{ \frac{n_1^2+n_2^2}{(n_1-n_2)^2}  +(1-a) }~.
\>
Expressing the form factor \eqref{eqn:ff-npw-Ls} in terms of $\Lc$, and including the rescaled normalisation \eqref{eqn:npw-norm-factor}, 
\< \label{eqn:ff-npw-Lc}
\hat{f}(n_1,n_2) \eq \frac{1}{2\Lc} + \frac{i\pi}{4L^2_c} \lrsbrk{ \frac{(n_1+n_2)^2}{n_1-n_2} + (1-2a) (n_1-n_2) } +\frac{1}{2\Lc^2} \left(\frac{n_1^2+n_2^2}{(n_1-n_2)^2}+(1-a)\right)  \nl
-\frac{\lambda}{8\Lc^3} (n_1^2+n_2^2) 
+ \frac{\sqrt{\lambda}}{2\Lc^3} \biggsbrk{ n_1^2+n_2^2 - a (n_1-n_2)^2 }   + \order(\Lc^{-4}) ~.
\>
Here, as in \eqref{eqn:ff-npw-Ls}, in the second line we have kept the subleading $1/L^3_c$ terms. In the usual BMN scaling the parameter $\tilde \lambda=\lambda/\Lc^2$ is taken to be small at both weak and strong 't Hooft coupling and the $1/\Lc^3$ terms should thus rather be interpreted as $\lt/\Lc$ and $\sqrt{\lt}/\Lc^2$ terms, respectively. In making a comparison with the tree-level gauge theory results we will not further consider these terms.

%%%%%%%%%%%%%%%%%%%%%%%%%%%%%%%%%%%%%%%%%%%%%%%%%%%%%%%
%%%%%%%%%%%%%%%%%%%%%%%%%%%%%%%%%%%%%%%%%%%%%%%%%%%%%%%
\subsection{World-Sheet versus Spin-Chain Operator}

The suitably rescaled form factor in spin-chain variables, \eqref{eqn:ff-npw-Lc}, is at leading order  similar to the spin-chain form factor \eqref{eq:spin_chain_ff}. However, it remains to carefully match the spin-chain operator to the world-sheet operator. 
 
\paragraph{Splitting the operator} In particular, in order to compare the finite volume form factor result to the spin-chain calculation at subleading order, we need take into consideration that the two $Y$-fields in \eqref{eqn:YY-operator} sit at the same point while the spin-chain operators $S_{+}$ in \eqref{eqn:spinchain-ff-def} act on distinct sites. We account for this difference by starting with the world-sheet operator $Y(x)Y(x+b)$ and Taylor expanding the second operator about $x$. This yields
\<
  \bra{0} Y(x)Y(x+b) \ket{p_1,p_2} = \bra{0} Y(x)Y(x) \ket{p_1,p_2} + \Half b^\mu \partial_\mu \bra{0} Y(x)Y(x) \ket{p_1,p_2} + \order(b^2)
\>
with $b^{\mu} = (0, b_{\mathrm{s}})$. In momentum space, the form factor for separated fields is thus
\<
  f_{\mathrm{sep}}(p_1,p_2) = f(p_1,p_2) - \iHalf b_{\mathrm{s}} (p_1+p_2) f(p_1,p_2) ~.
\>
On the spin-chain side, the two operators are separated by $b_c = 1$ sites. This should correspond to
\<
  b_{\mathrm{s}} = \frac{2\pi}{\sqrt{\lambda}}
\>
according to \eqref{eqn:length-conversion}. Using \eqref{eqn:string-momenta-modes-Lc} for the momenta, we have
\< \label{eqn:ws-corr-due-to-shift}
  - \iHalf b_{\mathrm{s}} (p_1+p_2) = - \frac{i\pi}{\Lc} (n_1+n_2) ~.
\>
This term must be added to the form factor, however it can be seen that it only contributes to the imaginary
term corresponding to the phase of the state. 

\paragraph{Operator normalization} A second issue is the exact map between the spin operators $S_+$ and the world-sheet fields $Y$. This  was discussed in context of world-sheet form factors in \cite{Klose:2012ju}. In general the relation is non-linear and at next-to-leading order is (e.g. see \appref{app:LL})
\<
S_+=\sqrt{2}Y\lrsbrk{1-\frac{3}{2}|Y|^2+\dots}~.
\>
For the two-particle form factors of the operator \eqref{eqn:YY-operator} we do not need to take into account the non-linear terms, but according to this mapping, the world-sheet operator $\half Y^2$ is a factor of $4$ smaller than the spin-chain operator $(S_{+})^2$.

\paragraph{Final result} 
Combining all the above factors and specialising to $a=1$ gauge we find for the operator ${\cal O}=2Y^2$, to order $1/\Lc^2$ and $\lt^0$
\< \label{eqn:final-result}
\hat{f}(n_1,n_2)=\frac{2}{\Lc}+\frac{2i \pi}{\Lc^2}\left(\frac{2 n_1n_2}{n_1-n_2}-(n_1+n_2)\right)
+\frac{2}{\Lc^2}\frac{n_1^2+n_2^2}{(n_1-n_2)^2}~.
\>
Compared to the spin-chain result  \eqref{eq:spin_chain_ff} we see that the real terms match while the imaginary terms do not. This is not too surprising as the phases of the states cannot be fixed. In principle we could include an additional overall phase factor in the definition of the spin-chain state that would give agreement (see \appref{app:alt_phase}). However, this means that the term coming from the bubble integral ($\sim 4i\pi$) cannot be compared to the spin-chain result. The first non-trivial comparison that can be performed is between the two-loop diagrams on the string side and the next-to-next-to-leading $1/\Lc$ term on the spin-chain side. Such terms could also be relevant when extending  the match between the string theory and gauge theory beyond leading order in the effective 't Hooft coupling $\lt$ and certainly such phases will be important in finding all-loop order solutions to the functional equations.

%%%%%%%%%%%%%%%%%%%%%%%%%%%%%%%%%%%%%%%%%%%%%%%%%%%%%%%%%%%%%%%%%%%%%%%%%%%
%%%%%%%%%%%%%%%%%%%%%%%%%%%%%%%%%%%%%%%%%%%%%%%%%%%%%%%%%%%%%%%%%%%%%%%%%%%
\section{Outlook}

In this work we have  continued earlier perturbative calculations of world-sheet form factors, \cite{Klose:2012ju}, by calculating the two-particle form factor for the $\algSU(2)$ sector operator ${\cal O}\sim Y^2$ to one loop in the near-plane-wave limit and to two loops in the Maldacena-Swanson limit. These perturbative calculations could of course be yet further continued to higher orders. At the level of explicit Feynman calculations one would expect the combinatorial complexity and the difficulties in performing the loop integrations to become increasingly burdensome and more efficient methods of calculation will be useful. The tools of generalised unitarity have recently been successfully applied to the calculation of the world-sheet S-matrix \cite{Bianchi:2013nra, Engelund:2013fja}. These methods are obviously analogous in part to the form factor axioms, i.e. they make use of the branch cuts and singularity structure, and it would be very interesting to apply them to the calculation of world-sheet form factors. 

While further perturbative calculations would be useful, the problem of finding exact solutions to the two-particle form factor axioms 
immediately presents itself. The generalised rapidity for the world-sheet magnons, $z$, is defined on a torus
with imaginary period $2\omega_2$ and the functional equation is written in terms of the exact world-sheet S-matrix $\Smat(z_1,z_2)$,
\<
f(z_1+2\omega_2,z_2)=S(z_1,z_2)f(z_1,z_2)~.
\> 
As for the relativistic case, one can write a formal solution as an infinite product
\<
f_{\rm min}(z_1,z_2)=\prod_{n=1}^\infty\Smat(z_1-2n\omega_2, z_2)~.
\>
however in this case, as we do not currently have a useful integral expression for the exact S-matrix, we cannot immediately write down a concrete expression following from this formal ``minimal" solution. Finding a generalisation of the relativistic contour argument or an analogous method will be a necessary step in determining the exact two-particle form factor. 

Extending these results beyond a simple rank-one sector to the full  world-sheet theory with $ \algPSU(2|2)^2\ltimes \mathbb{R}^3$ symmetry at the level of perturbative calculations should be straightforward. A more complete understanding via the axiomatic approach  will require significantly more powerful tools due to the non-diagonal scattering which results in matrix equations for the form factors and hence a more complicated algebraic structure. One approach to similar problems, for example in theories with $\algSU(N)$ factorised scattering, is the nested ``off-shell" Bethe ansatz \cite{Babujian:1990ii, Babujian:1993tm, Babujian:1993ts} as applied to form factors in, e.g., \cite{Babujian:2006md}. Another method for solving the form factor axioms, following ideas in the work \cite{Lukyanov:1992sc}, is based on finding the free field representation for the Zamolodchikov-Faddeev algebra \cite{Lukyanov:1993pn}. This method has been applied to a number of different models, see e.g. \cite{Horvath:1994rg, Lukyanov:1997bp, Alekseev:2009ik}; one model of interest in the current context is the chiral Gross-Neveu model with $\algSU(N)$ symmetry \cite{Britton:2013uta}. 

The world-sheet form factors become substantially more interesting quantities once we can understand their relation to observables in the ${\cal N}=4$ SYM. As described, they can be related to tree-level gauge theory structure constants via their match in the thermodynamic limit to spin-chain matrix elements. Recently, particularly following the work \cite{Escobedo:2010xs}, there has been a great deal of activity in extending the spin-chain methods in the calculation of structure constants \cite{Escobedo:2011xw,Gromov:2011jh,Foda:2011rr, Ahn:2012uv, Gromov:2012vu, Foda:2012wf,Serban:2012dr, Kostov:2012yq,Wheeler:2012bu, Gromov:2012uv, Kostov:2012wv, Foda:2012wn, Bissi:2012vx, Foda:2013gb, Serban:2013jua, Foda:2013nua, Kazama:2013rya, Wheeler:2013zja}, which may allow for a further study of the relation to world-sheet form factors. More generally, one may argue for a relation via the identification of gauge theory three-point correlation functions with world-sheet correlation functions of string vertex operators
\<
\langle {\cal O}_1(a_1) {\cal O}_2(a_2) {\cal O}_3(a_3)\rangle_{\rm CFT}\simeq \langle { V}_1(a_1){ V}_2(a_2) { V}_3(a_3)\rangle_{\rm world-sheet}~,
\>
where in our considerations ${ V}_i(a_i)$ is a world-sheet vertex operator dual to  a single trace gauge theory operator ${\cal O}_i(a_i)$ at a space-time point $a_i$. We  focus on the case where two of the string vertex operators, say $V_1$ and $V_3$, create near-BMN strings, that is strings with large energy and angular momentum, $J_1\simeq J_3\simeq J\sim \sqrt{\lambda}$ on an S$^2\in$ S$^5$ and some finite number of small momentum excitations. 
In light-cone gauge, the string vertex operators can be viewed as creating string world-sheets with specific excitations at world-sheet time $\tau_i\rightarrow \pm \infty$. If the remaining vertex operator $V_2$ creates a light string, i.e. one whose charges are $\leq \lambda^{1/4}$ we may attempt to treat it as a local operator on the world-sheet created (annihilated) by $ V_1$ ($V_3$) up to an overall factor capturing the dependence on the boundary location by assuming that it does not affect the semiclassical trajectory. In this heavy-heavy-light limit, the gauge theory structure constants should be related to the finite volume world-sheet matrix element. It would certainly be interesting to see to what degree this construction can be implemented. One possibility is to study the heavy-heavy-light limit for world-sheet correlation functions calculated in \cite{Dobashi:2002ar,Yoneya:2003mu,Dobashi:2004nm,Lee:2004cq,Shimada:2004sw} using methods based on plane-wave light-cone string field theory \cite{Spradlin:2003xc, Russo:2004kr} or by functional light-cone methods \cite{Klose:2011rm}. Given the possible relation to the heavy-heavy-light correlation functions, it would also be interesting to consider the relation of semi-classical form factors to the calculations of \cite{Zarembo:2010rr, Costa:2010rz}.

%

%%%%%%%%%%%%%%%%%%%%%%%%%%%%%%%%%%%%%%%%%%%%%%%%%%%%%%%%%%%%%%%%%%%%%%%%%%%
%%%%%%%%%%%%%%%%%%%%%%%%%%%%%%%%%%%%%%%%%%%%%%%%%%%%%%%%%%%%%%%%%%%%%%%%%%%
\section*{Acknowledgments}

We would like to thank Sergey Frolov for many useful discussions and comments. The work of TMcL was supported in part by Marie Curie Grant CIG-333851.  

\appendix

%

%%%%%%%%%%%%%%%%%%%%%%%%%%%%%%%%%%%%%%%%%%%%%%%%%%%%
%%%%%%%%%%%%%%%%%%%%%%%%%%%%%%%%%%%%%%%%%%%%%%%%%%%%%
\section{Operator Renormalization}
\label{app:Opren}

In this appendix we summarise a few standard facts about operator renormalization that are necessary for the calculation of the form factor. In particular, we are interested in the Green functions of the composite operator \eqref{eqn:YY-operator}, $\mathcal{O} = \half Y^2$, with fundamental fields
\<
G^{(n;1)}(\vec{p_1},\dots,\vec{p_n};\vec{q})&=&\langle\bar Y(\vec{p_1})\dots \bar Y(\vec{p_n}) \tilde{\cal{O}}(\vec{q})\rangle~,\nn\\
&=&\delta^{(2)}(\vec{q}+\sum_{i=1}^n\vec{p}_i)\prod_{i=1}^n\frac{i}{\vec{p}_i^2-1+i\eps}\hat{G}^{(n;1)}(\vec{p_1},\dots,\vec{p_n};\vec{q})~.
\>
At tree-level we have 
\<
\hat{G}_{tree}^{(2;1)}(\vec{p_1},\vec{p_2};\vec{q})=1
\>
while the one-loop result is 
\<
{\hat G}_{1-loop}^{(2;1)}(\vec{p_1},\vec{p_2};\vec{q})
&= &\left(\frac{2\pi i}{\sqrt{\lambda}}\right) \int \frac{d^d k}{(2\pi)^d}
\frac{(p_1+p_2)^2+(2a-1)(1-\vec{p_1}\cdot\vec{p_2}~ \vec{k}\cdot(\vec{q}-\vec{k}))}{(\vec{k}^2-1+i\eps )((\vec{q}-\vec{k})^2-1+i\eps)}\nn\\
& =&\left(\frac{-2\pi i}{\sqrt{\lambda}}\right) \Big[k^2 B_{00}(\vec{q})-(1-2a)(\vec{p_1}\cdot\vec{p_2} ~X(\vec{q})-B_{00}(\vec{q}))\Big]
\>
where we have regularised the loop integrations by working in $d=2-2\epsilon$ dimensions and the integrals are given to order
${\cal O}(\epsilon^0)$ by
\<
B_{00}(\vec{q})&=&-\frac{i}{\pi |\vec{q}|}\frac{1}{\sqrt{|\vec{q}|^2-4+4i\eps} }
\log \frac{ \sqrt{|\vec{q}|^2-4+4i\eps}+|\vec{q}| }{\sqrt{|\vec{q}|^2-4+4i\eps}+|\vec{q}|} \\
X(\vec{q})&=&\frac{i}{4\pi}\Big[\frac{1}{\epsilon}-\gamma_E+\log(4\pi)-\frac{2|\vec{q}|^2-4}{|\vec{q}|\sqrt{|\vec{q}|^2-4+4i\eps}}\log 
\frac{\sqrt{|\vec{q}|^2-4+4i\eps}+|\vec{q}|}{\sqrt{|\vec{q}|^2-4+4i\eps}+|\vec{q}|}\Big]
\>
for $\eps>0$ and $|\vec{q}|^2>4$. These are simply rewritings of the expressions used in the main text if one takes $\vec{q}=\vec{p_1}+\vec{p_2}$
with $\vec{p_1}$ and $\vec{p_2}$ on shell. In particular we write
\<
X(\vec{q})&=&\frac{i}{4\pi}C_\epsilon+X_R(\vec{q})
\>
with $C_\epsilon=\frac{1}{\epsilon}-\gamma_E+\log(4\pi)$.
We can now define a renormalized operator (in $\overline{\rm MS}$-scheme) as
\<
{{\cal O}}_R(\vec{x})=\frac{1}{2}Y^2-\left(\frac{2\pi i}{\sqrt{\lambda}}\right)\frac{(1-2a)}{2}\frac{i}{4\pi}C_\epsilon\partial_aY\partial^a Y
\>
such that
\<
{G}^{(2;1)}_R(\vec{p_1},\vec{p_2};\vec{q})&=&\langle\bar Y(\vec{p_1}) \bar Y(\vec{p_2}) \tilde{\cal{O}}_R(\vec{q})\rangle~,\nn\\
\Rightarrow \hat{G}^{(2;1)}_R(\vec{p_1},\vec{p_2};\vec{q}) &=&\left(\frac{-2\pi i}{\sqrt{\lambda}}\right) \Big[k^2 B_{00}(\vec{q})-(1-2a)(\vec{p_1}\cdot\vec{p_2} ~X_R(\vec{q})-B_{00}(\vec{q}))\Big]
\>
\paragraph{Two-loop two-point function}
It is also necessary to calculate the two-point world-sheet function of the composite (renormalized) composite operator
\<
G^{0;2}(\vec{q_1},\vec{q_2})&=&\langle \bar{{\tilde O}}(\vec{q_1})\tilde{ O}(\vec{q_2})\rangle~,
\>
which to  two-loops is 
\<
\label{eq:twopoint}
\hat{G}^{0;2}(\vec{q_1},\vec{q_2})
=-\frac{1}{2}B_{00}(\vec{q_1})
+\frac{1}{2}\left(\frac{2\pi i}{\sqrt{\lambda}}\right)
\Big[q_1^2 B_{00}(\vec{q_1})^2+(2a-1)(B_{00}(\vec{q_1})^2-X(\vec{q_1})^2)\Big]
\>
where we use $B_{00}(\vec{-q})=B_{00}(\vec{q})$ and $X_R(-\vec{q})=X_R(\vec{q})$. Using the counter terms found above, and removing an overall divergence proportional to the identity operator, we find that the renormalized Green function is found by simply making the replacement $X(\vec{q})\rightarrow X_R(\vec{q})$ in \eqref{eq:twopoint}.

%%%%%%%%%%%%%%%%%%%%%%%%%%%%%%%%%%%%%%%%%%%%%%%%%%%%%%%%%%%%%%%%%%%%%%%%%%%
%%%%%%%%%%%%%%%%%%%%%%%%%%%%%%%%%%%%%%%%%%%%%%%%%%%%%%%%%%%%%%%%%%%%%%%%%%%
\section{Determinant Form for Spin-Chain Norm }
\label{sec:sc-norm-from-Jacobian}

The mode numbers label the solutions of the Bethe equations, which in the two-magnon sector read
\<
  e^{ip_1\Lc} = e^{i\Theta_{\mathrm{c}}(p_1,p_2)}
	\comma
  e^{ip_2\Lc} = e^{-i\Theta_{\mathrm{c}}(p_1,p_2)}
\>
with $\Theta_{\mathrm{c}}(p_1,p_2)$ given in \eqref{eqn:phase-shift}. Taking the logarithm of these equations,
\< \label{eqn:Bethe-log}
  p_1 \Lc = \Theta_{\mathrm{c}}(p_1,p_2) + 2\pi n_1
	\comma 
  p_2 \Lc = -\Theta_{\mathrm{c}}(p_1,p_2) + 2\pi n_2
	~.
\>
By differentiating each of the equations in \eqref{eqn:Bethe-log} by $n_1$, we produce two equations that can be solved for the partial derivatives:
\<
  \frac{\partial p_1}{\partial n_1} \eq \frac{2\pi}{\Lc} \lrbrk{ 1 + \frac{2 (1-\cos p_2)}{3 \Lc-4 - 2 (\Lc-1)( \cos p_1 +\cos p_2) + \Lc\cos(p_1+p_2)}} ~, \\
  \frac{\partial p_2}{\partial n_1} \eq -\frac{2\pi}{\Lc} \frac{2(1-\cos p_2 )}{3 \Lc-4 - 2 (\Lc-1)( \cos p_1 +\cos p_2) + \Lc\cos(p_1+p_2)} ~.
\>
Similarly we can differentiate the equations with respect to $n_2$ yielding expressions for the partial $n_2$-derivatives, corresponding to the above expressions with $p_1\leftrightarrow p_2$ everywhere (also on the left hand side). Now, we can compute the Jacobian and find
\< \label{eqn:Jacobian-eval}
  \lrabs{\frac{\partial( p_1, p_2 )}{\partial(n_1,n_2)}}^{-1} =
	\frac{\Lc}{4\pi^2} \lrbrk{ \Lc - \frac{2 (2 - \cos p_1 - \cos p_2)}{3 - 2 ( \cos p_1 + \cos p_2 ) + \cos(p_1+p_2)} }~.
\>
This agrees with \eqref{eqn:sc-scalar-product-exact} up to an overall factor of $4\pi^2$.

%%%%%%%%%%%%%%%%%%%%%%%%%%%%%%%%%%%%%%%%%%%%%%%%%%%%%%%%%%%%%%%%%%%%%%%%%%%
%%%%%%%%%%%%%%%%%%%%%%%%%%%%%%%%%%%%%%%%%%%%%%%%%%%%%%%%%%%%%%%%%%%%%%%%%%%
\section{Alternative Spin-Chain Wave-Function}
\label{app:alt_phase}

As explained in the main text, see below \eqref{eqn:final-result}, the imaginary, subleading $\frac{1}{L_c}$ terms in the spin-chain form factor depend on the choice of the overall phase of the wave-function. In this appendix, we present a phase that yields a match with the world-sheet form factor, namely
\<
  \chi(p_1,p_2)_{x_1,x_2} = e^{-2i(p_1+p_2)} \Bigsbrk{ e^{ip_1x_1 + ip_2x_2} + S(p_2,p_1) e^{ip_2x_1 + ip_1x_2} }~,
\>
instead of \eqref{eqn:two-spin-wave-function}. The normalization factor $\mathcal{N}_{\mathrm{c}}(p_1,p_2)$ is not changed by this additional phase factor and its large $\Lc$ limit is given by \eqref{eqn:sc-norm-factor}. Expanding also the above wave-function for large $\Lc$ using \eqref{eqn:spin-chain-momenta-modes} gives
\<
  \chi(p_1,p_2)_{1,2} = 2 - \frac{2\pi i}{\Lc} \frac{n_1^2 - 2n_1n_2 - n_2^2}{n_1-n_2} + \order(\Lc^{-2}) ~.
\>
Putting everything together, the form factor becomes to subleading order in $1/\Lc$
\<
  f(p_1,p_2) = \frac{2}{\Lc} - \frac{2\pi i}{\Lc^2} \frac{n_1^2 - 2n_1n_2 - n_2^2}{n_1-n_2} + \frac{2}{\Lc^2} \frac{n_1^2 + n_2^2}{(n_1-n_2)^2} + \order(\Lc^{-3}) ~,
\>
which matches \eqref{eqn:final-result} including the imaginary terms.

%%%%%%%%%%%%%%%%%%%%%%%%%%%%%%%%%%%%%%%%%%%%%%%%%%%%
%%%%%%%%%%%%%%%%%%%%%%%%%%%%%%%%%%%%%%%%%%%%%%%%%%%%%
\section{Comparison between Landau-Lifshitz and String Actions}
\label{app:LL}

In order to better understand the matching between the string theory calculation 
and the spin-chain calculation it is useful to reconsider the match that is found
at the level of the actions via the Landau-Lifshitz action\cite{Kruczenski:2003gt, Kruczenski:2004kw}. The spin-chain
in the thermodynamic limit can be described by a unit vector field $\vec{n}(\sigma, \tau)$ with the action
\<
S_{LL}=\frac{{\Lc}}{2\pi}\int d\tau \int^{2\pi}_0 d\sigma \Big[\vec{C}\cdot \partial_\tau\vec{n}-\frac{1}{8} \lt (\partial_\sigma\vec{n})^2\Big]
\>
where ${\Lc}$ is the spin-chain vacuum length and $\lt=\lambda/{\Lc}^2$
\footnote{In fact one must be slightly careful here. The usual definition of the rescaled
't Hooft coupling in the literature is $\lt=\lambda/J^2$ where 
$J$ is the R-charge of the operator whose anomalous dimension is under 
consideration. Obviously the relation between $\Lc$ and $J$ varies
depending on the number of impurities in the spin-chain state. However 
this won't affect our considerations.}
 $\vec{C}$ is a monopole potential on S$^2$ such
that the action can be written locally (where $n_3\neq -1$) as 
\<
S_{LL}=\frac{{\Lc}}{4\pi}\int d\tau \int^{2\pi}_0 d\sigma \Big[\frac{n_2\partial_\tau n_1 - n_1\partial_\tau n_2}{1+n_3}-\frac{1}{4} \lt (\partial_\sigma\vec{n})^2\Big]~,
\>
where $n_3=\sqrt{1-n_1^2-n_2^2}$.
This action can be quantised, and with the appropriate regularization, loop corrections reproduce the subleading $1/{\Lc}$ corrections
to the exact spin-chain energies. Introducing a complex field $\phi=\tfrac{1}{2}(n_1+i n_2)$ we find 
\<
S_{LL}=\frac{{\Lc}}{4\pi}\int d\tau \int^{2\pi}_0 d\sigma \Big[\frac{i(\phi^\ast \dot \phi-\phi\dot \phi^\ast)}{1+\sqrt{1-4|\phi|^2}}+\lt|\acute \phi|^2+\lt \frac{(\phi^\ast \acute\phi+\phi\acute\phi^\ast)^2}{1-4|\phi|^2}\Big]~.
\>
Rescaling the spatial coordinate $\sigma\rightarrow \sqrt{\lt} \sigma$, rescaling the fields by $\sqrt{{\Lc}/2\pi}$ and expanding in large ${\Lc}$ we find, 
\<
S_{LL}=\frac{1}{2} \int d\tau \int^{\frac{2\pi}{\sqrt{\lt}} }_0 d\sigma \Big[  i(\phi^\ast \dot \phi-\phi\dot \phi^\ast)- |\acute \phi|^2+ \frac{2\pi i}{{\Lc}}|\phi|^2 (\phi^\ast \dot \phi-\phi\dot \phi^\ast)-
\frac{2\pi }{{\Lc}} (\phi^\ast \acute\phi+\phi\acute\phi^\ast)^2\Big]~.
\>
We note that the world-sheet length is $\Ls=\frac{2\pi}{\sqrt{\lt}}=\frac{2\pi {\Lc}}{\sqrt{\lambda}}$ while the loop counting parameter is $\frac{2\pi }{{\Lc}}$.

In the main text we calculated the perturbative form factors using 
the action defined on the plane, i.e. the decompactified world-sheet, using the string theory action in general $a$-gauge. 
However in order to make comparison with the spin-chain calculation we need to consider the theory on the cylinder, here of length $2\pi$, and we 
will also use the $a=1$ gauge
\<
S=\frac{\sqrt{\lambda}}{2\pi}\int d\tau \int^{2\pi}_{0} d\sigma \Big[   \partial Y \partial \bar{Y} - Y \bar{Y}
+ 2 Y \acute{Y} \bar{Y} \acute{\bar{Y}} - \frac{1}{2} \Bigbrk{ (\partial Y)^2 (\partial \bar{Y})^2 -  Y^2 \bar{Y}^2 }\Big]~.
\>
We now introduce the new coordinate $Y=y e^{-i\tau}$,  rescale the world-sheet time $\tau\rightarrow \kappa \tau$, the
world-sheet spatial coordinate $\sigma\rightarrow \sqrt{\kappa} \sigma$ and expand in large $\kappa$ keeping only the leading
term, 
\<
S=\frac{\sqrt{\lambda \kappa}}{2\pi}\int d\tau \int^{\frac{2\pi}{\sqrt{\kappa}}}_{0} d\sigma &\Big[&  \kern-7pt i (y^\ast \dot y-y\dot y^\ast)-|\acute y|^2+2 |y|^2 | \acute y|\nn\\
& &-\frac{1}{2}\left(2 i |y|^2 (y^\ast \dot y - y\dot y^\ast)+y^\ast{}^2 \acute y^2+y^2\acute y^\ast{}^2 \right) \Big]~.
\>
Making the substitution $y=\sqrt{\frac{\pi}{\sqrt{\lambda \kappa}}}\phi(1+\frac{3}{4}\frac{2\pi}{\sqrt{\lambda \kappa}}|\phi|^2)$ expanding in large $\sqrt{\lambda\kappa}$ we find 
\<
S=\frac{1}{2} \int d\tau \int^{\frac{2\pi}{\sqrt{\kappa}} }_0 d\sigma \Big[  i(\phi^\ast \dot \phi-\phi\dot \phi^\ast)- |\acute \phi|^2+ \frac{2\pi i}{\sqrt{\lambda \kappa}}|\phi|^2 (\phi^\ast \dot \phi-\phi\dot \phi^\ast)-
\frac{2\pi }{\sqrt{\lambda \kappa}} (\phi^\ast \acute\phi+\phi\acute\phi^\ast)^2\Big]~.
\>
Quite obviously,  to find agreement with the LL action we make the identification $\lambda \kappa={\Lc}^2$ i.e. $\kappa=\lt^{-1}$. We can also 
see that having calculated the perturbative form factors with the string theory in the decompactified theory we can obtain the spin-chain
answer by recompactifying the world-sheet with length $\Ls=2\pi/\lt^{1/2}$ and replacing the loop counting parameter $\frac{1}{\sqrt{\lambda}}$ 
by $\frac{1}{{\Lc}}$ and expanding the answer in small $\lt$ (naturally this will hold as long as there are no order of limits issues but this is the 
underlying assumption behind the weak-strong match which has been found to hold to this perturbative order). 
\pdfbookmark[1]{\refname}{references}
\bibliographystyle{nb}
\bibliography{FormFactors}

\end{document}